\documentstyle[12pt,epsfig,subfigure,amsfonts]{article}

\newcommand{\be}{\begin{equation}}
\newcommand{\ee}{\end{equation}}
\newcommand{\beq}{\begin{equation}}
\newcommand{\eeq}{\end{equation}}
\newcommand{\ba}{\begin{eqnarray}}
\newcommand{\ea}{\end{eqnarray}}
\newcommand{\bea}{\begin{eqnarray}}
\newcommand{\eea}{\end{eqnarray}}

\begin{document}
\baselineskip=15.5pt \pagestyle{plain} \setcounter{page}{1}
\begin{titlepage}

\vskip 0.8cm

\begin{center}

{\LARGE Thermalization with a chemical potential from AdS spaces}
\vskip .3cm

\vskip 1.cm

{\large {Dami\'an Galante{\footnote{\tt dagalante@gmail.com}} and
Martin Schvellinger{\footnote{\tt martin@fisica.unlp.edu.ar}}}}

\vskip 1.cm

{\it IFLP-CCT-La Plata, CONICET and \\
Departamento  de F\'{\i}sica, Universidad Nacional de La Plata.
\\ Calle 49 y 115, C.C. 67, (1900) La Plata,  \\ Buenos Aires,
Argentina.} \\

\vspace{1.cm}

{\bf Abstract}

\end{center}

The time-scale of thermalization in holographic dual models with a
chemical potential in diverse number of dimensions is systematically
investigated using the gauge/gravity duality. We consider a model
with a thin-shell of charged dust collapsing from the boundary
toward the bulk interior of asymptotically anti-de Sitter (AdS)
spaces. In the outer region there is a Reissner-Nordstr\"om-AdS
black hole (RNAdS-BH), while in the inner region there is an anti-de
Sitter space. We consider renormalized geodesic lengths and minimal
area surfaces as probes of thermalization, which in the dual quantum
field theory (QFT) correspond to two-point functions and expectation
values of Wilson loops, respectively. We show how the behavior of
these extensive probes changes for charged black holes in comparison
with Schwarzschild-AdS black holes (AdS-BH), for different values of
the black hole mass and charge. The full range of values of the
chemical potential over temperature ratio in the dual QFT is
investigated. In all cases, the structure of the thermalization
curves shares similar features with those obtained from the AdS-BH.
On the other hand, there is an important difference in comparison
with the AdS-BH: the thermalization times obtained from the
renormalized geodesic lengths and the minimal area surfaces are
larger for the RNAdS-BH, and they increase as the black hole charge
increases.

\noindent

\end{titlepage}

\newpage

\tableofcontents

\vfill

\newpage

\section{Introduction}

Quark gluon plasma (QGP) has become a subject of considerable
interest after the results from RHIC and LHC experiments of heavy
ion collisions, from where it is possible to draw the conclusion
that the system behaves like an ideal fluid with a small shear
viscosity over entropy density ratio ($\eta/s$)
\cite{Shuryak:2003xe,Schukraft:2011kc,Abreu:2007kv}. The crucial
outcome is that the QGP is a strongly coupled system. Theoretical
calculations assume that the system is in thermal equilibrium, and
it is probed at momentum scales below the equilibrium temperature
$T$, {\it i.e.} in the hydrodynamical regime of the plasma. A
suitable approach to the dynamical description of strongly coupled
systems is based on the AdS/CFT correspondence
\cite{Maldacena:1997re,Gubser:1998bc,Witten:1998qj}. In this
context, many investigations have been carried out after the seminal
paper by Policastro, Son and Starinets, where they proposed a
relation between the shear viscosity of the finite-temperature
${\cal {N}} = 4$ supersymmetric $SU(N)$ Yang-Mills (SYM) theory
plasma, in the large $N$ limit, at strong-coupling regime, and the
absorption cross section of low-energy gravitons by a near-extremal
black three-brane \cite{Policastro:2001yc}. The ${\cal {N}} = 4$ SYM
theory at finite temperature (just above the deconfinement
transition temperature $T_c$) in the large $N$ limit, qualitatively
behaves like the strongly coupled QGP. This statement is based on a
number of lattice QFT computations (for a recent review see
\cite{CasalderreySolana:2011us}). The results of
\cite{Policastro:2001yc} were obtained within the supergravity
limit. The first leading order string theory corrections to $\eta/s$
of strongly coupled SYM plasmas have been computed in
\cite{Buchel:2004di}. On the other hand, electrical charge transport
properties of strongly coupled SYM plasmas have been thoroughly
studied as well. These include the electrical conductivity, which
was firstly computed in \cite{Policastro:2002se} within the
supergravity limit, while its string theory corrections were
obtained in \cite{Hassanain:2010fv,Hassanain:2011fn}. The first
computation of photoemission rates in this context was done in
\cite{CaronHuot:2006te} and their corresponding leading order string
theory corrections were obtained in \cite{Hassanain:2011ce}.

The way in which the actual heavy ion collision process is
understood involves four basic stages. It begins with the {\it
approach} regime, where two heavy ions move toward each other at
relativistic velocities, with kinetic energies typically about 100
GeV/nucleon. Soon after, once the heavy ions collide, a part of
their kinetic energies transforms into intense heat and the plasma
of quarks and gluons starts forming. This is the {\it
thermalization} process. After that, the strongly coupled QGP is
formed, and here is where the hydrodynamical description holds.
Elliptic flow has been observed in this regime, and it has been
concluded that the system behaves like an ideal fluid, {\it i.e.}
with very low $\eta/s$. At the end, the system expands and cools
down, leading to the {\it hadronization} process. Since QGP in
thermal equilibrium is strongly coupled, in principle, it is
possible that the thermalization described above also occurs within
a strongly coupled regime of QCD. Regardless of whether or not this
hypothesis holds, it is very interesting to think about the
thermalization of a strongly coupled system. In fact, in a series of
papers the issue has been studied considering the collapse of a thin
shell of matter, by using an AdS-Vaidya type metric
\cite{Das:2010yw,AbajoArrastia:2010yt,Albash:2010mv,Ebrahim:2010ra,
Balasubramanian:2010ce,Balasubramanian:2011ur,Garfinkle:2011hm,
Aparicio:2011zy,Allais:2011ys,Keranen:2011xs,
Garfinkle:2011tc,Das:2011nk,Hubeny:2012ry}. Earlier studies of
holographic thermalization described as a dual process of black hole
formation have been presented in references
\cite{Danielsson:1999fa,Giddings:1999zu,Janik:2006gp,Janik:2006ft,
Chesler:2008hg,Chesler:2009cy,Bhattacharyya:2009uu,Lin:2008rw}.

In a more realistic physical situation one should also consider the
chemical potential, $\mu$. Thus, it is very interesting to study the
effect of the chemical potential on the thermalization time scale.
This can be qualitatively modeled by a collapsing thin shell of
charged matter, leading to a thermal equilibrium configuration given
by a Reissner-Nordstr\"om-AdS black hole. In this paper we carry out
a detailed systematic investigation, exploring the full range of the
chemical potential/equilibrium temperature ratio, $\mu/T$, in
diverse number of dimensions, for both renormalized geodesic lengths
and minimal area surfaces. In a real QGP the ratio $\mu/T$ is about
$0.15$ or less, thus the chemical potential effects will be limited
\cite{Adams:2005dq,Back:2004je,Adcox:2004mh,Arsene:2004fa,Myers:2009ij}.
For instance, at supergravity level, $\eta/s$ remains the same with
finite chemical potential. However, it is interesting to consider
the effect of the chemical potential on the thermalization time
scale for different holographic probes, mainly to be able to
investigate the behavior of a collapsing shell of charged matter in
AdS spaces. This may have applications to other physical systems
beyond the context of SYM plasmas, such as certain condensed matter
systems in two and three dimensions. We would like to emphasize that
although the main motivation for the present work has been to
consider thermalization of QGPs, which concerns RNAdS$_5$ black
holes, the systematic exploration extended to lower and higher
dimensions presented here would perhaps motivate further studies
from the AdS/CMT perspective (for 3 and 4 dimensional backgrounds)
as well as higher dimensional dual QFTs (corresponding to 6 and 7
dimensional backgrounds). It certainly will be of interest to
investigate whether these effective backgrounds can be lifted to
ten-dimensional string theory or eleven-dimensional M-theory
descriptions, or at least to understand the difficulties to do it.

In section 2 we describe a model of holographic thermalization at
strong coupling using two types of extensive geometric probes which
are dual to two-point functions of QFT gauge invariant local
operators and expectation values of rectangular Wilson loops. In
section 3 we will describe the Reissner-Nordstr\"om-AdS black hole
in arbitrary number of dimensions. There is an important reason to
carry out a systematic study in different dimensions, which is the
fact that the probes of thermalization such as renormalized geodesic
lengths and minimal area surfaces are one- and two-dimensional
geometric objects, respectively. Thus, by increasing the number of
background spacetime dimensions, it allows to explore the ability of
these extended objects to probe additional bulk degrees of freedom.
This effect is significant as we will show from the figures
introduced in the last section of the paper. We will introduce a
time-dependent AdS-Vaidya type metric, including charged matter in
the collapsing shell. This is essential for the present study of
thermalization, and with it we will study thermalization of
two-point functions of gauge invariant local operators in the dual
conformal field theory (CFT), as well as rectangular Wilson loops,
in the whole range $0 \leq \mu/T \le \infty$. This is done for QFTs
in 3, 4, 5 and 6 dimensions. In the case of AdS$_3$/CFT$_2$ duality
there are charged extensions of the BTZ black hole
\cite{Martinez:1999qi}, however these are special cases that we
shall not discuss here. Our results show that the effect of
introducing chemical potential produces a delay in the
thermalization time of the dual strongly coupled system. This effect
indeed becomes more evident as the size of geometric probe
increases. Our results are consistent with a top-down thermalization
process, which was already observed in previous computations with no
chemical potential
\cite{Balasubramanian:2010ce,Ebrahim:2010ra,Balasubramanian:2011ur},
and is inherent to a strongly coupled computation. Thus, as it
happens for the AdS-BH case, in the RNAdS-BH geometry UV modes
thermalize first. This is a natural consequence of the setup of a
collapsing thin shell, since smaller geodesics and smaller minimal
area surfaces dual to Wilson loops thermalize before the
corresponding larger probes do. It would be interesting to consider
fat collapsing shells and see whether the top-down thermalization
persists. Another observation from this systematic analysis is that,
for sufficiently large probes, a swallow-tail like pattern appears
in the thermalization curves, both for renormalized geodesic lengths
and minimal area surfaces. This is a known effect in systems with
two different relevant scales. We shall discuss about the relevance
of these scales. On the other hand, this pattern could possibly be a
consequence of a numerical artifact due to the fact that numerically
we never have a shell with zero thickness. In addition, the swallow
tails could appear from the breakdown to the probe approximation
near the horizon, as it happens in top-down models when considering
extremal probes instead of fundamental string probes including the
thermalization of the string degrees of freedom
\cite{Grignani:2012iw}. After a detailed numerical analysis our
conclusion is that the swallow tails are due to the emergence of
certain solutions, from which only one is relevant for
thermalization.

It is interesting to note the relation between two-point functions
and expectation values of Wilson loops with the entanglement entropy
in AdS$_3$ and AdS$_4$, respectively. Ryu and Takayanagi
\cite{Ryu:2006bv,Ryu:2006ef,Nishioka:2009un} proposed a way to
calculate entanglement entropy holographically by considering
minimal area surfaces in AdS spaces. Thus, by computing geodesics in
AdS$_3$ and minimal area surfaces in AdS$_4$, we will also be
computing the evolution of the entanglement entropy during the
thermalization of CFTs in 2 and 3 dimensions, respectively, at
finite temperature and finite chemical potential.

Finally, it is also interesting to note the differences in
comparison with previous recent works. For a more complete
description of investigations related to scenarios of thermalization
of an initial field configuration far from equilibrium see
\cite{Balasubramanian:2011ur} and references therein. In
\cite{Balasubramanian:2010ce,Balasubramanian:2011ur} it is
considered the thermalization from a collapsing shell of
pressureless dust leading to Schwarzschild-AdS black holes in 3, 4
and 5 dimensions. The thermalization probes in these references are
renormalized geodesic lengths, minimal area surfaces and minimal
volumes. Then in \cite{Albash:2010mv} it is studied the evolution
and scaling of the entanglement entropy for a 3 dimensional field
theory using a thermal quench, leading to an asymptotically anti-de
Sitter Schwarzschild black hole in four dimensions. In addition,
they consider an electromagnetic quench, which after complete
thermalization, renders an extremal dyonic black hole in four
dimensions. In this reference the entanglement entropy is studied.
More recently, in reference \cite{Keranen:2011xs} thermalization
following a non-relativistic quench near a quantum critical point
with non-trivial dynamical critical exponent has been studied. They
consider a collapsing shell leading to asymptotically Lifshitz
spaces. In our present work we consider all cases for AdS$_{d+1}$
spaces with $d=$3, 4, 5, and 6, exploring the full range of $\mu/T$,
considering different values of the boundary separation $\ell$. We
also study the swallow tails emerging from the thermalization
curves, for renormalized geodesic lengths and minimal area surfaces.
We also show explicitly the shape of geodesics corresponding to zero
and finite chemical potential, for a given boundary separation
$\ell$, considering a set of different times. This allows to show
graphically the actual shape evolution of the geodesics as the thin
shell moves toward the bulk interior.

\section{A model of holographic thermalization at strong coupling}

In this section we describe a model of holographic thermalization of
a strongly coupled CFT based on
\cite{Balasubramanian:2010ce,Balasubramanian:2011ur}. We present a
Vaidya-type metric where a thin shell of charged pressureless dust
collapses to form a charged, asymptotically anti-de Sitter black
hole. Then, we analyze two types of QFT operators: two-point
functions of local gauge invariant operators and expectation values
of rectangular Wilson loops, that will be used in the next sections
as probes of holographic thermalization.

We will generalize the Vaidya metric, which was originally
introduced by P.C. Vaidya in 1943 in order to study how the geometry
of spacetime evolves in the presence of a massive star, which varies
its mass due to the effect of radiation \cite{vaidya1,vaidya2}. In
the present case, we consider $d+1$-dimensional metrics that are
asymptotically AdS$_{d+1}$. Thus
\begin{eqnarray}
ds^2=\frac{1}{z^2} \left[ - f(v,z) dv^2 - 2 dz\ dv +
d\textbf{x}^2 \right] \, , \label{metrica_vaidya}
\end{eqnarray}
where $f(v,z)$ can be an arbitrary function of $v$ and $z$ which
approaches 1 as $z$ goes to zero. The AdS radius, $R$, has been set
to 1. The $z$-coordinate is the inverse of the radial coordinate,
$z=R^2/r$, so the AdS boundary is at $z=0$. In addition, \textbf{x}
stands for coordinates at the boundary except the time,
$\textbf{x}=(x_1,\cdots,x_{d-1})$. In particular, if we choose
$f(v,z)= 1 - m(v) z^d$ we will be considering a shell (with no
charge) that collapses to form a Schwarzschild-AdS black hole.
Moreover, we can extend this geometry to include charge by defining
\begin{eqnarray}
f_{RN}(z,v) = 1- m(v) z^d + q(v)^2 z^{2d-2} \, , \label{f_RN}
\end{eqnarray}
where the subindex stands for Reissner-Nordstr\"om black hole and
the expression holds for $d\geq 3$.

This type of metric is a solution of the equations of motion of the
action
\begin{eqnarray}
S_{RN-Vaidya}=S_{EMAdS}+S_{ext},
\end{eqnarray}
where $S_{EMAdS}$ is the Einstein-Maxwell anti-de Sitter action
\begin{eqnarray}
S_{EMAdS} = -\frac{1}{16\pi G^{(d+1)}} \int d^{d+1}x \sqrt{-g}
\left[{\cal{R}} - F^2 + d (d-1) \right] \, , \label{accion RN 1}
\end{eqnarray}
where the negative cosmological constant is defined as $\Lambda = -
\frac{d (d-1)}{2}$ and $S_{ext}$ is an external action, whose
energy-momentum tensor and current will be specified below.

From this action the resulting equations of motion are
\begin{eqnarray}
R_{\mu\nu}-\frac{1}{2} g_{\mu\nu} ({\cal{R}}-2 \Lambda - F^2) - 2 F_{\mu \lambda} F_{\nu}^{\lambda} & = & 8\pi G^{(d+1)} T_{\mu\nu}^{(ext)} \, , \\
\frac{1}{\sqrt{-g}} \partial_{\rho} \left( \sqrt{-g} F^{\rho \sigma}
\right) & = & 8\pi G^{(d+1)} J^{\sigma}_{(ext)} \, .
\end{eqnarray}

The AdS-Vaidya type metric presented in Eqs.(\ref{metrica_vaidya})
and (\ref{f_RN}), together with an electromagnetic potential defined
as
\begin{eqnarray}
A_\mu = - \frac{1}{c} q(v) z^{d-2} \delta_{\mu v},
\end{eqnarray}
with $c=\sqrt{\frac{2 (d-2)}{d-1}}$, satisfy the above equations of
motion provided that the external currents are
\begin{eqnarray}
8 \pi G^{(d+1)} T_{\mu\nu}^{(ext)} & = & z^{d-1} \left( \frac{(d-1)}{2} \dot{m} (v)
- (d-1) z^{d-2} q(v) \dot{q} (v) \right) \delta_{\mu v} \delta_{\nu v} \, , \\
8 \pi G^{(d+1)} J_{(ext)}^{\sigma} & = & \sqrt{\frac{(d-1)(d-2)}{2}}
z^{d+1} \dot{q} (v) \delta^{\sigma z} \, ,
\end{eqnarray}
where the dot stands for derivative with respect to coordinate $v$.
Note that in the chargeless case, the energy-momentum tensor only
depends on $\dot{m} (v)$ so it is immediate to check that the null
energy conditions are satisfied provided that $m(v)$ is
non-decreasing function of $v$. The charged case is more subtle
since it depends on the specific choice of both $m(v)$ and $q(v)$.

The functions $m(v)$ and $q(v)$ model the change in the mass and
charge of the black hole. In particular, if both $m(v)=M$ and
$q(v)=Q$ are constants, then $v$ is the usual Eddington-Finkelstein
coordinate,
\begin{eqnarray}
dv=dt-\frac{dz}{1-M \ z^d + Q^2 z^{2d-2}} \, , \label{cv_dv}
\end{eqnarray}
where $t$ is the time. If we rewrite the metric in terms of $t$ and
$z$, then the AdS-Vaidya type metric corresponds to a charged
black-brane type geometry,
\begin{eqnarray}
ds^2=\frac{1}{z^2} \left[ - (1- M z^d + Q^2 z^{2d-2}) dt^2
+ \frac{dz^2}{1-M \ z^d + Q^2 z^{2d-2}}
+ d\textbf{x}^2 \right] \, . \label{metrica_ads_bh}
\end{eqnarray}
Again, for $Q=0$, we obtain a black-brane solution with an event
horizon at $r_h=1/z_h=M^{1/d}$.

On the other hand, if $m(v)=q(v)=0$ the change of variables becomes
$dv=dt-dz$, being the metric the usual AdS$_{d+1}$ one. Thus, if we
choose $m(v)=M \, \theta(v)$ and $q(v)=Q \, \theta(v)$, with
$\theta$ being the step function, we obtain a spacetime separated
into two regions by a shell: for $v<0$ it is an AdS$_{d+1}$
spacetime, otherwise it will be a Reissner-Nordstr\"om-AdS black
hole. At $v=0$ we have a shell of zero thickness. Since we will
carry out a numerical study it is convenient to consider smooth
functions like
\begin{eqnarray}
m(v) & = & \frac{M}{2} \left( 1 + \tanh \frac{v}{v_0} \right) \, , \label{masa rn vaidya} \\
q(v) & = & \frac{Q}{2} \left( 1 + \tanh \frac{v}{v_0} \right), \label{carga rn vaidya}
\end{eqnarray}
where $v_0$ represents a finite shell thickness. In the limit $v_0
\rightarrow 0$ we recover the  step function that represents a shock
wave. In this limit case the null energy conditions are also
satisfied, while in the finite thickness case the analysis has to be
done with some care (see Appendix for a complete discussion on
this).

Once we have a metric to model the rapid injection of energy that
initiates the thermalization process on the dual QFT, we have to
choose a set of extended observables in the bulk which allow to
evaluate the evolution of the system. The initial condition is that
the spacetime is AdS$_{d+1}$. Then, after certain time elapses, the
system reaches the thermodynamical equilibrium, and the bulk becomes
a Reissner-Nordstr\"om-AdS black hole. A RNAdS-Vaidya type metric
interpolates between these two spaces, describing the time evolution
of the system. In order to follow this evolution we have to probe
the bulk by using extended geometric objects, which should be
computed before and after the system reaches the thermodynamical
equilibrium. However, it is easy to imagine that not any observable
can be used as a probe of thermalization. One may ask what happens
with the expectation values of one-point functions of gauge
invariant operators, such as the energy-momentum tensor for
instance. The problem is that these are local operators and,
therefore, cannot give information on the thermalization process.
So, clearly one needs to consider extended probes. A possibility is
to study correlation functions of two local gauge invariant
operators on the field theory side.

The AdS/CFT correspondence allows to understand intuitively why the
expectation values of local quantum field theory operators are
insensible to the details of the thermalization. These observables
can only account for effects due to bulk properties close to the AdS
boundary, thus they do not allow to probe details from distances of
the order of the thermal scale of the system. On the other hand,
two-point functions of local gauge invariant QFT operators probe the
bulk interior. Indeed, the AdS/CFT correspondence provides an
elegant geometric way to compute two-point functions: under certain
approximations the two-point functions correspond to geodesics which
connect the two points where the local QFT operators are inserted at
the AdS boundary. These geodesics extend toward the bulk, thus
allowing to probe a larger range of energies from the boundary field
theory perspective. There are also other operators which are useful
for the present problem: expectation values of Wilson loops which
correspond to minimal area surfaces in the AdS bulk, as well as the
entanglement entropy which is related to minimal volumes.

Following \cite{Balasubramanian:2011ur} we first consider two-point
functions of local gauge invariant QFT operators at constant time as
probes of thermalization. In general Wightman functions are defined
as \cite{Balasubramanian:2010ce,Balasubramanian:2011ur}
\begin{equation}
G_{{\cal{O}}}^> (t,\textbf{x}; t', \textbf{x}') = < {\cal{O}}
(t,\textbf{x}) {\cal{O}}(t', \textbf{x}')> \, , \label{eqG}
\end{equation}
where ${\cal{O}}$ is a local gauge invariant QFT operator of
conformal dimension $\Delta$. In order to follow the evolution of
these functions we are interested in the equal time ($t=t'$)
correlators. Then, we want to see how these functions vary at
different times. When we consider strongly coupled QFTs these
functions can only be computed analytically for dimension two. For
higher dimensions the analytical treatment is unknown.

Fortunately, the AdS/CFT correspondence allows to evaluate these
functions when the operators are heavy by using geodesics in AdS
spaces. As it is well-known from the AdS/CFT dictionary a scalar
field $\varphi (z, t, \textbf{x})$ with mass $m$ in $d+1$ dimensions
is dual to an operator ${\cal{O}}$ whose conformal dimension is
$\Delta=\frac{1}{2} (d+\sqrt{d^2+m^2})$. In general, the two-point
function (\ref{eqG}) in the strongly coupled regime is computed
using the classical supergravity action in terms of $\varphi$.
However, for our purpose it will be more convenient to compute it
from a path integral as in reference
\cite{Balasubramanian:1999zv,Balasubramanian:2011ur}
\begin{eqnarray}
< {\cal{O}} (t,\textbf{x}) {\cal{O}}(t, \textbf{x}')> = \int
{\cal{D}}{\cal{P}} e^{i \Delta L({\cal{P}})} \approx \sum_{
\textrm{geodesics}} e^{-\Delta {\cal{L}}} \, ,
\end{eqnarray}
where the path integral includes all possible paths connecting the
points at the AdS boundary, {\it i.e.} $(t, \textbf{x})$ and $(t,
\textbf{x}')$. In addition, $L({\cal{P}})$ is the proper length
corresponding to this path. For space-like trajectories
$L({\cal{P}})$ is imaginary. Thus, it is possible to make a
saddle-point approximation for $\Delta \gg 1$. Therefore, only
trajectories with extreme geodesic lengths will contribute. Notice
that in the last term ${\cal{L}}$ indicates actual length of the
geodesic between the points at the AdS boundary. In this way, there
is a direct relation between the logarithm of the equal-time
two-point function and the geodesic length between these two points.
It is important to be careful while considering these approximations
because the geodesic length diverges due to the AdS boundary
contributions. Then, one can define a renormalized distance $\delta
{\cal{L}} \equiv {\cal{L}} - 2 \ln (2/z_0)$, in terms of the cut-off
$z_0$. This suppresses the divergent part in the pure AdS.

The other type of non-local operators that we will be using are
spatial Wilson loops, which are non-local gauge invariant operators
in the field theory defined as the integral in a closed path
\textit{C} of the gauge field
\begin{eqnarray}
W(C) = \frac{1}{N} Tr \left( {\cal{P}} e^{\oint_C A} \right) \, ,
\end{eqnarray}
where ${\cal{P}}$ means ordered product, $N$ is the number of colors
of the gauge theory, and $A_\mu$ is the non-Abelian gauge field.
Wilson loops provide information about the non-perturbative behavior
of gauge theories, however, in general it is difficult to compute
them. Using the AdS/CFT correspondence its computation can be done
in an elegant way. The expectation value of a Wilson loop is related
to the string theory partition function with a world sheet $\Sigma$
extended on the interior of the bulk, and ending on the closed
contour \textit{C} on the boundary,
\begin{eqnarray}
<W(C)> = \int {\cal{D}}\Sigma e^{-\Lambda(\Sigma)} \, ,
\end{eqnarray}
where one has to integrate over all the non-equivalent surfaces
whose boundary is  $\partial\Sigma=C$, at the AdS boundary.
$\Lambda(\Sigma)$ is the string action. In the strong coupling
regime we can carry out a saddle-point approximation of the string
theory partition function. In this way we can reduce the computation
of the expectation value of a Wilson loop to determine the surface
of minimal area of the classical world-sheet whose boundary is
\textit{C}. Thus, can write
\begin{eqnarray}
<W(C)> \simeq e^{-\frac{1}{\alpha'}{\cal{A}}(\Sigma_0)} \, ,
\end{eqnarray}
where ${\cal{A}}(\Sigma_0)$ represents the area of the minimal
surface. This will be a solution to the equations of motion of the
bosonic part of the string action \cite{malda 2,rey}.

These models based on the AdS/CFT correspondence allow to understand
intuitively how the thermalization process takes place. Let us
consider the sudden injection of energy in the dual QFT. The bulk
geometry associated with this process is proposed to be described by
a collapsing shell from the boundary toward the bulk interior. As
long as the shell collapses, the outer region is described by a
Reissner-Nordstr\"om-AdS black hole, while the inner region is still
an AdS space. Now, let us use the geodesic approximation to compute
the equal-time two-point functions. If the separation of the
boundary points is small enough, then the geodesic cannot reach the
shell at $v=0$ and, therefore, the geodesic is seen as a purely
RNAdS-BH geodesic, {\it i.e.} for short distances in the field
theory the system seems to be in thermal equilibrium. If we increase
the separation between the insertion of the boundary operators, at
some point, the geodesic will cross the shell, and there will be a
geodesic refraction which will deviate the geodesic in comparison
with the thermal one. Thus, we can understand why the thermalization
proceeds from short to long distances, {\it i.e.} QFT ultraviolet
degrees of freedom thermalize first \cite{Balasubramanian:2011ur}.

\section{The Reissner-Nordstr\"om-AdS black hole}

Let us now consider the pure Einstein-Maxwell anti-de Sitter action
(with no external fields) which describes a $d+1$-dimensional
space-time, with negative cosmological constant $\Lambda=-\frac{d
(d-1)}{2R^2}$, coupled to an Abelian gauge field $A_M$, with $M=0,
\cdot \cdot \cdot, d$ \cite{Chamblin:1999tk,Chamblin:1999hg}
\begin{eqnarray}
S_{EMAdS} = -\frac{1}{16\pi G^{d+1}} \int d^{d+1}x \sqrt{-g}
\left[{\cal{R}} - F^2 + \frac{d (d-1)}{R^2} \right] \, .
\label{accion RN}
\end{eqnarray}
This action has been introduced in Section 2, however here we
recover the AdS radius $R$. The solution to the equations of motion
derived from the above action leads to the Reissner-Nordstr\"om-AdS
black hole metric, which can be written in static coordinates as
\begin{eqnarray}
ds^2 = - V(r) dt^2 + \frac{dr^2}{V(r)} + r^2 d\Omega_{d-1}^2 \, ,
\end{eqnarray}
where $d\Omega_{d-1}^2$ is the metric on the sphere $S^{d-1}$, and
\begin{eqnarray}
V(r) = 1 + \frac{r^2}{R^2} - \frac{M}{r^{d-2}} +
\frac{Q^2}{r^{2d-4}} \, . \label{u de erre 0}
\end{eqnarray}
In this metric $M$ and $Q$ are related to the ADM black hole mass
$\bar{M}$ and the $\bar{Q}$ charge \cite{Emparan:1998pf} as follows
\begin{eqnarray}
\bar{M} & = & \frac{(d-1) \omega_{d-1}}{16 \pi G} M \, , \\
\bar{Q} & = & \sqrt{2 (d-1)(d-2)} \left( \frac{\omega_{d-1}}{8 \pi
G}\right) Q \, ,
\end{eqnarray}
where $\omega_{d-1}$ is the volume of the unit radius sphere
$S^{d-1}$. Also, there is a pure electrical gauge potential given by
\begin{eqnarray}
A = \left( -\frac{1}{c} \frac{Q}{r^{d-2}} + \Phi\right) dt \, ,
\end{eqnarray}
where $c=\sqrt{\frac{2 (d-2)}{d-1}}$, while $\Phi$ is a constant
which plays the role of the electrostatic potential in the region
between the event horizon and the boundary of the asymptotic AdS. It
is defined such that $A_t (r_h) = 0$. Thus,
\begin{eqnarray}
\Phi = \frac{1}{c} \frac{Q}{r_h^{d-2}} \, .
\end{eqnarray}

Following \cite{Chamblin:1999tk} we can consider the limit where the
boundary of AdS$_{d+1}$ is $\mathbb{R}$${}^d$ instead of
$\mathbb{R}$ $\times S^{d-1}$, the so-called infinite volume limit,
which is important for the dual field theory discussion. This relays
upon the presence of a negative cosmological constant. This limit
can be obtained by introducing a dimensionless parameter $\lambda$,
and set for the radial coordinate $r \rightarrow \lambda^{1/d} r$,
for time coordinate $t \rightarrow \lambda^{-1/d} t$, for the mass
and charge $m \rightarrow \lambda m$ and $q \rightarrow
\lambda^{(d-1)/d} q$, respectively, and for $S^{d-1}$ we set $R^2
d\Omega_{d-1}^2 \rightarrow \lambda^{-2/d} \sum_{i=1}^{d-1} dx_i^2$.
Then, if we consider the limit $\lambda \rightarrow \infty$, the
metric becomes
\begin{eqnarray}
ds^2 = - U(r) dt^2 + \frac{dr^2}{U(r)} + \frac{r^2}{R^2}
\sum_{i=1}^{d-1} dx_i^2 \, , \label{metricr}
\end{eqnarray}
where
\begin{eqnarray}
U(r) = \frac{r^2}{R^2} - \frac{M}{r^{d-2}} + \frac{Q^2}{r^{2d-4}} \,
. \label{u de erre}
\end{eqnarray}
Notice that replacing $r$ by $1/z$ (setting $R=1$) in
Eqs.(\ref{metricr}) and (\ref{u de erre}) we obtain
Eq.(\ref{metrica_ads_bh}).

Using the gauge/gravity duality dictionary the Hawking temperature
corresponding to a black hole is assumed to be the equilibrium
temperature of the dual QFT. For the RNAdS-BH metric the temperature
is given by
\begin{eqnarray}
\beta_{RN} = \frac{1}{T_{RN}} = \frac{4\pi R^2}{d r_h- \frac{(d-2)
Q^2 R^2}{r_h^{2d-3}}} \, , \label{rn temp}
\end{eqnarray}
in terms of the black hole event horizon $r_h$ and its charge
parameter $Q$. This expression results from the usual procedure of
demanding regularity at the event horizon of the Euclidean
continuation of the metric. When  $Q=0$ this equation leads to
$\beta = \frac{4\pi R^2}{d r_h}$, being the well-known Hawking
temperature in the AdS-BH case. Then, for instance, setting $d=4$ it
reduces to the Hawking temperature of the Reissner-Nordstr\"om
AdS$_5$ black hole \cite{Myers:2009ij},
\begin{eqnarray}
\beta_{RN} = \frac{1}{T_{RN}}= \frac{\pi R^2}{r_h (1- \frac{Q^2
R^2}{2r_h^6})} \, .
\end{eqnarray}
We can also consider the black hole mass parameter in terms of the
radius of the event horizon
\begin{eqnarray}
M = \frac{r_h^d}{R^2} + \frac{Q^2}{r_h^{d-2}} \, . \label{masa RN}
\end{eqnarray}

We should notice that the RNAdS black hole has an extremal solution
corresponding to $T=0$. However, it nevertheless has an event
horizon. This can be seen from Eq.(\ref{rn temp}), leading to
\begin{eqnarray}
\left. r_h \right|_{ext} = \left( \frac{(d-2)}{d} Q^2 R^2
\right)^{\frac{1}{2d-2}} \, . \label{caso extremo}
\end{eqnarray}
This is a remarkable difference with respect to the
Schwarzschild-AdS case, where the extremal solution corresponds to
an anti-de Sitter spacetime. Another important difference is that
from the AdS/CFT correspondence it is possible to understand the
electromagnetic field in this geometry as a source of the chemical
potential in the dual quantum field theory. However, the precise
relation is somehow subtle since the chemical potential has energy
units in the dual field theory ($[\mu] = 1/[L]$)
\cite{Myers:2009ij}. On the other hand, $A_\mu$ as defined in the
action (\ref{accion RN}) is dimensionless, thus one has to redefine
this field as $\bar{A}_\mu = A_\mu / R_*$, where $R_*$ is a scale
with length units. Therefore, $\bar{A}$ and $\mu$ have the same
units. The only effect of this field redefinition is to change the
Maxwell term in the action, changing the coupling as $g_{YM}^2=4\pi
G/R_*^2$. This is useful now because in this way we can obtain the
chemical potential in the QFT as the boundary value of the time
component of the bulk gauge field. Thus, we obtain
\begin{eqnarray}
\mu = \lim_{r\rightarrow\infty} \bar{A}_t = \frac{\Phi}{R_*} =
\frac{1}{c} \frac{Q}{r_h^{d-2} R_*} \, . \label{mu}
\end{eqnarray}
Therefore, the black hole charge is related to the chemical
potential in the gauge theory. $R_*$ depends on the particular
compactification, whenever the solution could be obtained from
string theory or M-theory. Besides the specific value of $R_*$, it
is possible to study the whole range of values of $\mu/T$ given by
\begin{equation}
\frac{\mu}{T} = \frac{4 \pi \, R^2 \, Q}{c  \, R_* \left( d \,
r_h^{d-1} - \frac{(d-2) Q^2 R^2}{r_h^{d-1}}\right)} \, .
\end{equation}
Thus, if the radius of the event horizon $r_h$ is kept fixed, then
Eqs.(\ref{caso extremo}) and (\ref{mu}) show that it is possible to
go from $\mu/T=0$, for $Q=0$, to the extremal case, where $\mu/T
\rightarrow\infty$, provided that $Q$ satisfies  Eq.(\ref{caso
extremo}). Recall that the mass is given by Eq.(\ref{masa RN}).

\section{Holographic thermalization with a chemical potential}

In this section we consider how the thermalization process occurs
within the model described here. For this purpose we will study
two-point functions of local gauge invariant operators and
expectation values of rectangular Wilson loops in the dual conformal
field theory. The former ones correspond to geodesic lengths in the
bulk theory, while the second ones correspond to minimal area
surfaces in the bulk. The idea is that we start from an anti-de
Sitter spacetime of dimension $d+1$. This is the holographic dual of
the strongly coupled regime of a conformal field theory at zero
temperature and zero chemical potential. Then, we consider a thin
shell of charged dust propagating from the boundary toward the bulk
interior, and collapsing, leading to a RNAdS black hole. From the
dual QFT point of view this corresponds to a sudden injection of
energy and matter, both modeled through holographic quenches, such
that whenever the thermodynamical equilibrium is reached, the system
will be described within the grand canonical ensemble. As the thin
shell of charged matter collapses in the bulk, it separates two
regions. The outer one is a Reissner-Nordstr\"om black hole which is
asymptotically AdS. The inner region is just an AdS spacetime.

This situation is described by a dynamical metric, which is the
RNAdS Vaidya type metric given in Eqs.(\ref{metrica_vaidya}) and
(\ref{f_RN}), using the functional forms of the mass and charge of
Eqs.(\ref{masa rn vaidya}) and (\ref{carga rn vaidya}). The
conjecture is that from the holographic dual field theory this
dynamical situation corresponds to the thermalization process of a
strongly coupled system. We will see that the holographic
thermalization is a top-down process in the sense that UV degrees of
freedom equilibrate first.

Now, we can actually calculate the geodesic length and the minimal
area surface of the string attached to the boundary, as probes of
thermalization for this geometry.

\subsection{Renormalized geodesic lengths}

We will be evaluating geodesic distances as function of both time
and boundary separation length, in arbitrary number of spacetime
dimensions. We will consider space-like geodesics between points
$(t,x_1)=(t_0, -\ell/2)$ and $(t',x_1')=(t_0, \ell/2)$ in the case
of AdS$_3$/CFT$_2$, where $\ell$ is the separation of the AdS
boundary points. For $d=3, 4, 5, 6$, the orthogonal coordinates are
fixed. For instance, for $d=4$ we have $(x_2,x_3)=(x_2',x_3')$.
Thus, we take as the geodesic parameter the first coordinate $x_1$,
which, in order to make simpler the notation, we rename as $x$. The
solutions to the geodesic equations are given by the functions
$v(x)$ and $z(x)$. Inserting a cut-off $z_0$ close to the AdS
boundary, the boundary conditions become
\begin{equation}
z(-\ell/2)  =  z_0 \, , \,\,\,\,\,\,\,\, z(\ell/2)   =  z_0 \, ,
\,\,\,\,\,\,\,\, v(-\ell/2) = t_0 \, , \,\,\,\,\,\,\,\,  v(\ell/2) =
t_0 \, .
\end{equation}
Also, $v(x)$ and $z(x)$ are symmetric under reflection $x
\rightarrow -x$. The geodesic length is defined as
\begin{eqnarray}
{\cal{L}}  =  \int \sqrt{-ds^2} = \int_{-\ell/2}^{\ell/2} dx
\frac{\sqrt{1-2z'(x)v'(x) - f_{RN}(v,z) v'(x)^2}}{z(x)} \, ,
\label{ltermal}
\end{eqnarray}
where $f_{RN}(v,z)$ was defined in Eq.(\ref{f_RN}). The prime
indicates derivative with respect to $x$. The functions $v(x)$ and
$z(x)$ minimize the geodesic length of Eq.(\ref{ltermal}), thus the
problem is similar to the one in classical mechanics with the same
Lagrangian. Since the Lagrangian does not depend explicitly on $x$,
there is one conserved quantity, equivalent to the Hamiltonian of
the system. In terms of $f_{RN}(v,z)$ it becomes
\begin{equation}
{\cal{H}}  =  \frac{1}{z(x) \sqrt{1-2z'(x)v'(x) - f_{RN} (v,z)
v'(x)^2} } \, .
\end{equation}
Introducing the initial conditions on the tip of the geodesic as
\begin{eqnarray}
z(0)=z_* \, , \hspace{1.5cm} v(0)=v_* \, , \hspace{1.5cm} v'(0) =
z'(0) = 0 \, ,
\end{eqnarray}
the conservation equation simplifies to
\begin{eqnarray}
1- 2 z' v' - f_{RN}(z,v) v'^2 = \left( \frac{z_*}{z} \right)^2 \, .
\label{conservacion_two}
\end{eqnarray}

The next step is to compute the equations of motion for $v(x)$ and
$z(x)$. Though in principle they have a complex form, it is possible
to simplify them by making use of the conservation equation. In
particular, we can differentiate Eq.(\ref{conservacion_two}) with
respect to $x$ and obtain a relation for $v''(x)$ and $z''(x)$.
Inserting these relations in the equations of motion, we obtain a
set of simplified, second order differential equations for $v(x)$
and $z(x)$ as follows
\begin{eqnarray}
0 & = & 2 - 2 v'(x)^2 f_{RN}(v,z) - 4 v'(x) z'(x) - 2 z(x) v''(x) +
z(x) v'(x)^2 \partial_z f_{RN}(v,z) \, , \label{t_eom1}
\end{eqnarray}
which is obtained from the equation of motion for $z(x)$, while
\begin{eqnarray}
0 & = & z(x) v''(x) f_{RN}(v,z) + z(x) z''(x) + z(x) z'(x) v'(x)
\partial_z f_{RN}(v,z) \nonumber\\
& & + \frac{1}{2} z(x) v'(x)^2 \partial_v f_{RN}(v,z) \, ,
\label{t_eom2}
\end{eqnarray}
is obtained from the equation of motion of $v(x)$. From this set of
equations (plus the initial values) we can numerically obtain $z(x)$
and $v(x)$ for each pair of $v_*$ and $z_*$. We can also obtain the
equations of motion in terms of $m$ and $q$, just by replacing
$f_{RN}$ and its derivatives into Eqs.(\ref{t_eom1}) and
(\ref{t_eom2}), leading to
\begin{eqnarray}
0 & = &  \frac{1}{2} d \, z(x)^d  v'(x)^2 m(v) - z(x)^d v'(x)^2 m(v)
- d \, z(x)^{2 d-2} v'(x)^2 q(v)^2
\nonumber \\
&&  + 2 z(x)^{2 d-2} v'(x)^2
q(v)^2 + z(x) v''(x) + 2 v'(x) z'(x) + v'(x)^2 - 1 \label{de_z} \, , \\
0 & = & -\frac{1}{2} z(x)^d v'(x)^2 m'(v) - z(x)^d v''(x) m(v) - d
\, z(x)^{d-1} v'(x) z'(x) m(v) \nonumber \\
& & + z(x)^{2 d-2} v'(x)^2 q(v) q'(v) + z(x)^{2 d-2} v''(x) q(v)^2 +
2 d \, z(x)^{2 d-3} v'(x) z'(x) q(v)^2 \nonumber \\
&& - 2 z(x)^{2 d-3} v'(x) z'(x) q(v)^2 + v''(x) + z''(x) \, ,
\label{de_v}
\end{eqnarray}
where the dot stands for partial derivative with respect to $v$.

The advantage of calculating the equations of motion in terms of
$f_{RN}(v,z)$ is that they can be easily generalized to any other
metric with that form. We can also obtain the equations of motion
for the Schwarzschild-AdS case simply by changing $f_{RN}(v,z)$ for
$f(v,z)=1-m(v) z^d$, {\it i.e.} setting $q(v)=0$. It leads to the
equations for geodesic lengths presented in
\cite{Balasubramanian:2011ur}, which is a consistency check for our
calculations and it also allows to compare the RNAdS-BH with the
Schwarzschild-AdS one.

At the end we are interested in evaluating the geodesic length in
terms of time $t_0$ and the boundary separation $\ell$. This
information is given by
\begin{eqnarray}
z(\ell/2)=z_0 \, ,  \hspace{1.5cm} v(\ell/2)=t_0 \, .
\end{eqnarray}

By using the conservation equation and reflection symmetry we can
easily calculate the on-shell geodesic length as
\begin{eqnarray}
{\cal{L}} (\ell, t_0) = 2 \int_{0}^{\ell/2} dx \frac{z_*}{z(x)^2}
\, ,
\end{eqnarray}
and subtract the divergent part defining $\delta {\cal{L}} (\ell,
t_0) = {\cal{L}} (\ell, t_0) - 2 \ln (2/z_0)$.

Thus, we can calculate how the thermalization process occurs by
considering a collapsing thin shell of charged dust. It is
interesting to compare these results with those obtained in the case
of a collapsing thin shell of dust with no charge, such as those
described in \cite{Balasubramanian:2011ur}.

\begin{figure}
\centering
\subfigure[$t_0=0.498$ - AdS-Vaidya]{
\includegraphics[scale=0.45]{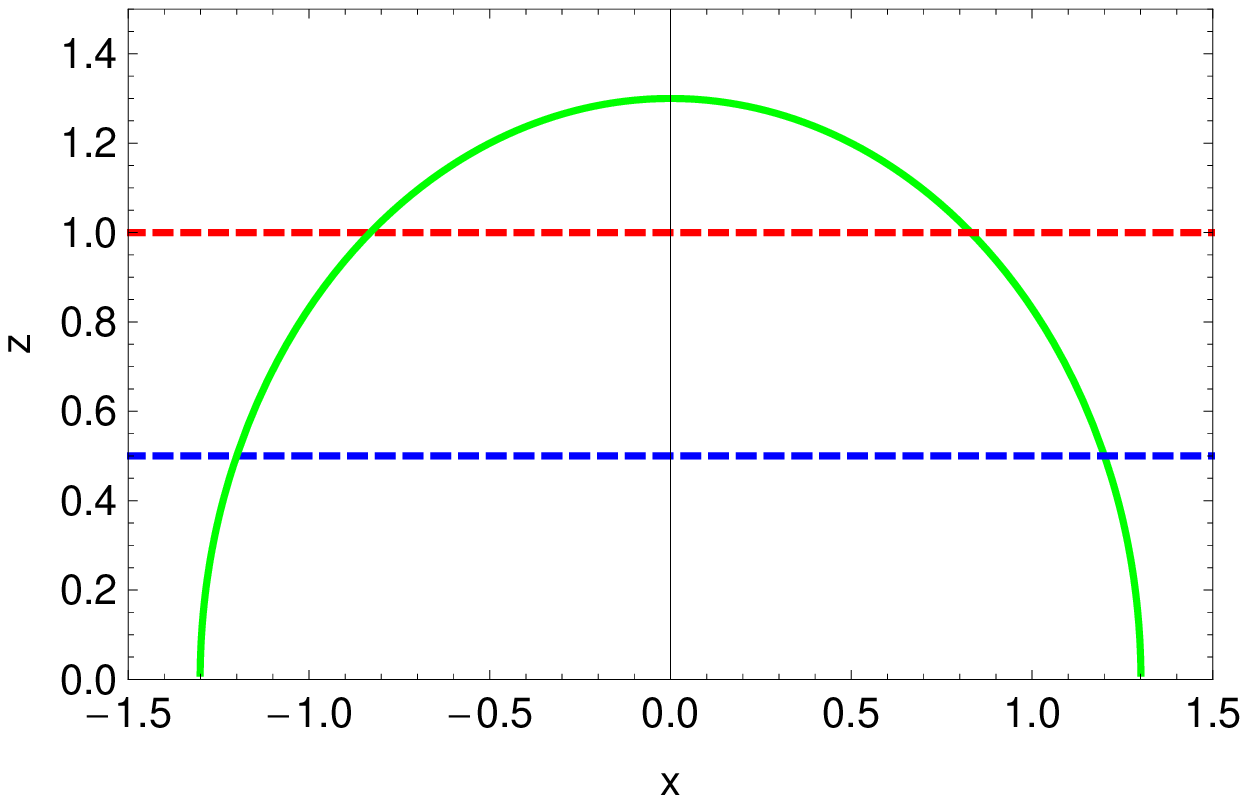}
\label{shell1_RN} } \subfigure[$t_0=0.498$ - RNAdS Vaidya]{
\includegraphics[scale=0.45]{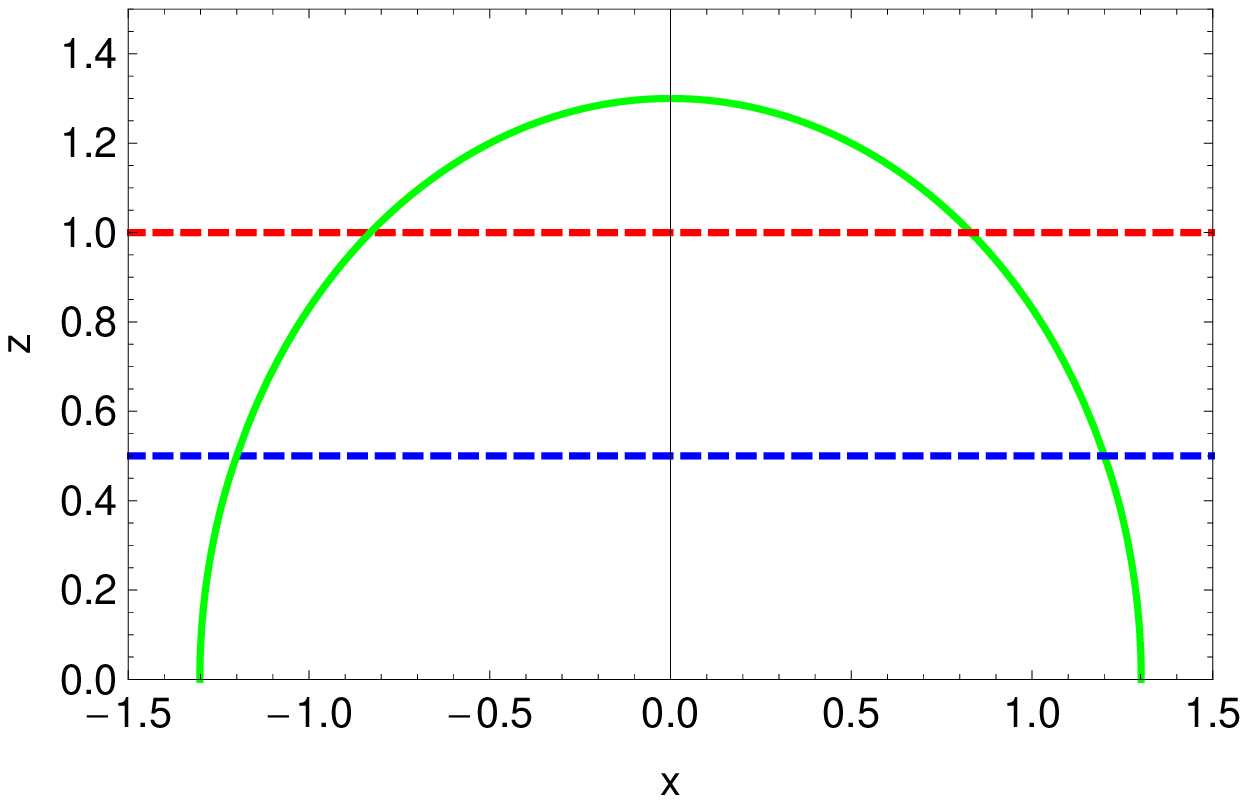}
\label{shell2_RN}
}
\subfigure[$t_0=0.778$ - AdS-Vaidya]{
\includegraphics[scale=0.45]{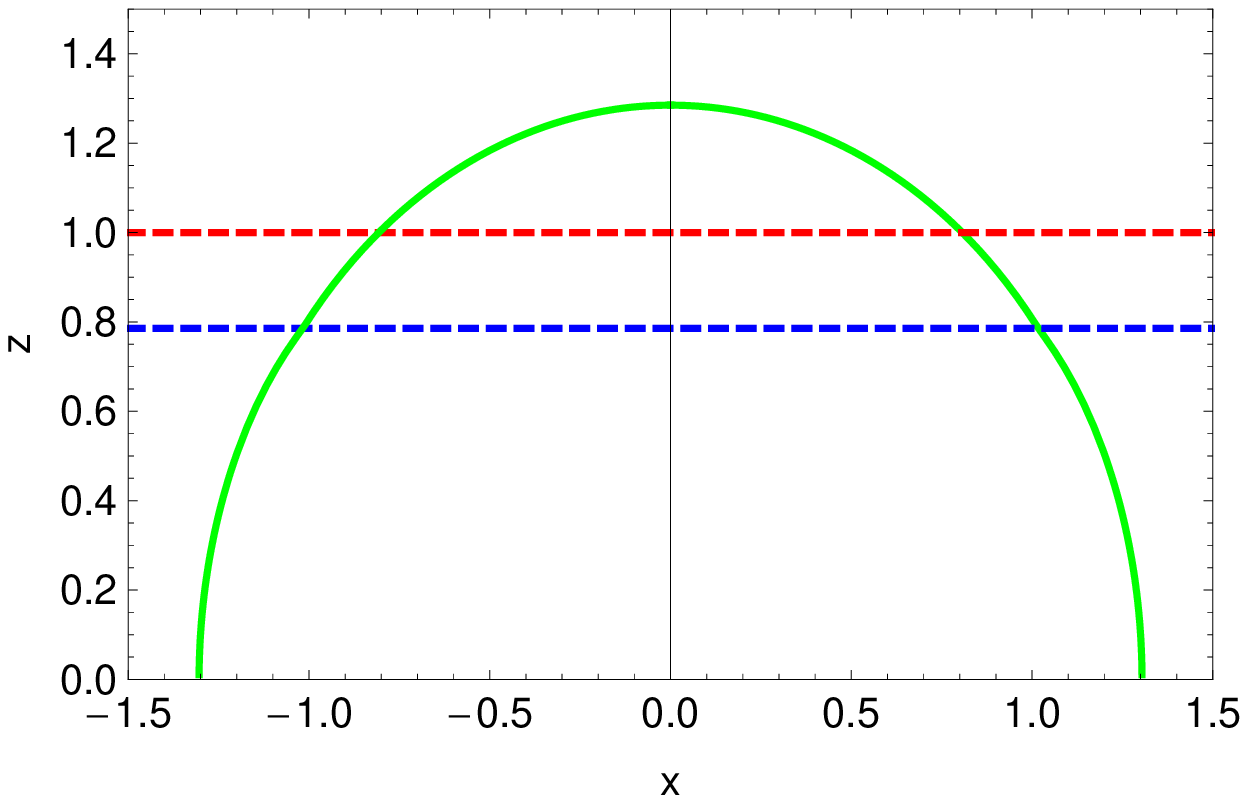}
\label{shell3_RN} } \subfigure[$t_0=0.771$ - RNAdS Vaidya]{
\includegraphics[scale=0.45]{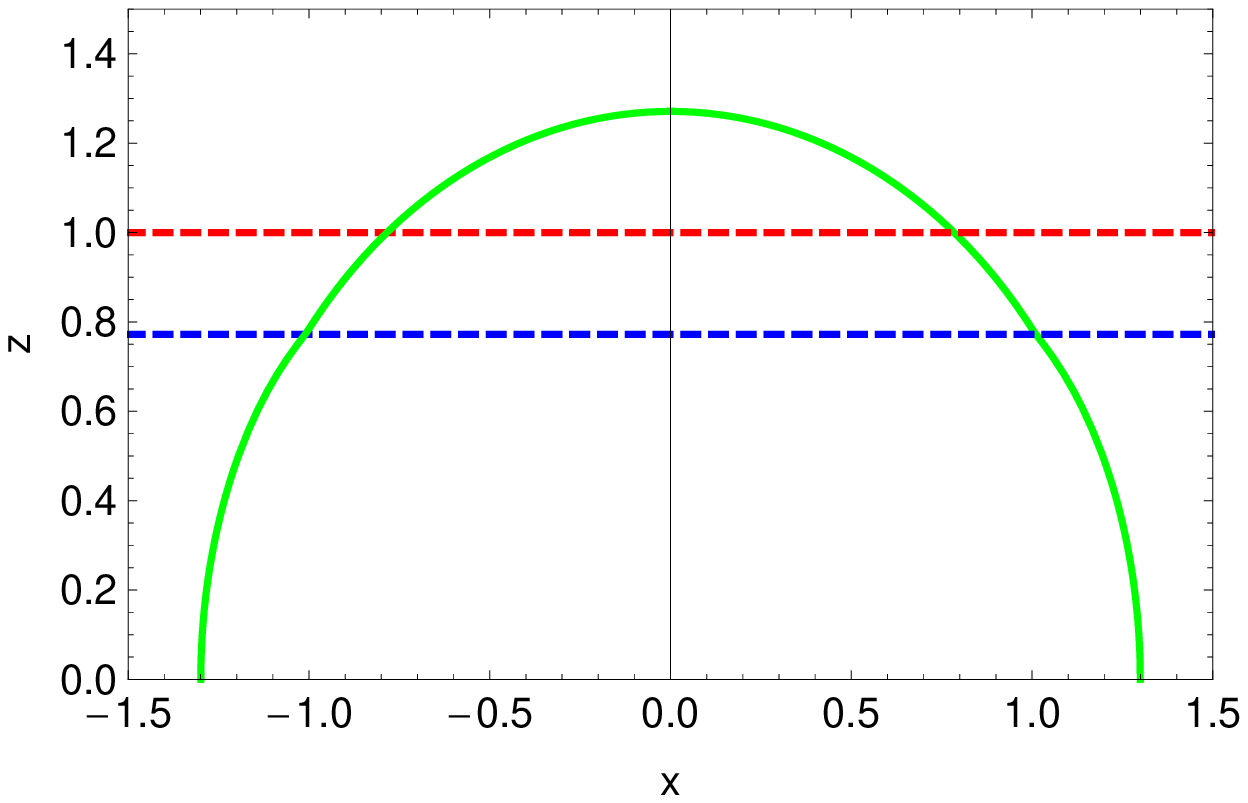}
\label{shell4_RN}
}
\subfigure[$t_0=1.010$ - AdS-Vaidya]{
\includegraphics[scale=0.45]{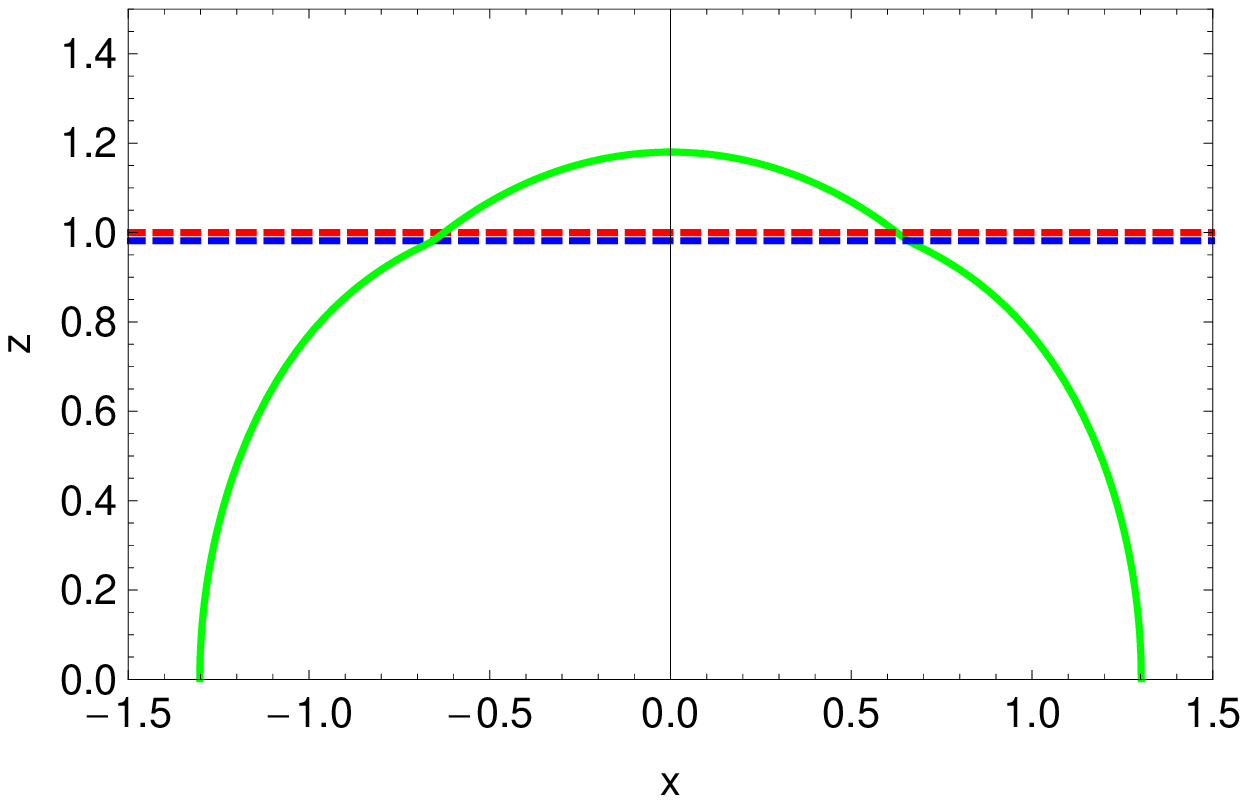}
\label{shell5_RN} } \subfigure[$t_0=1.011$ - RNAdS Vaidya]{
\includegraphics[scale=0.45]{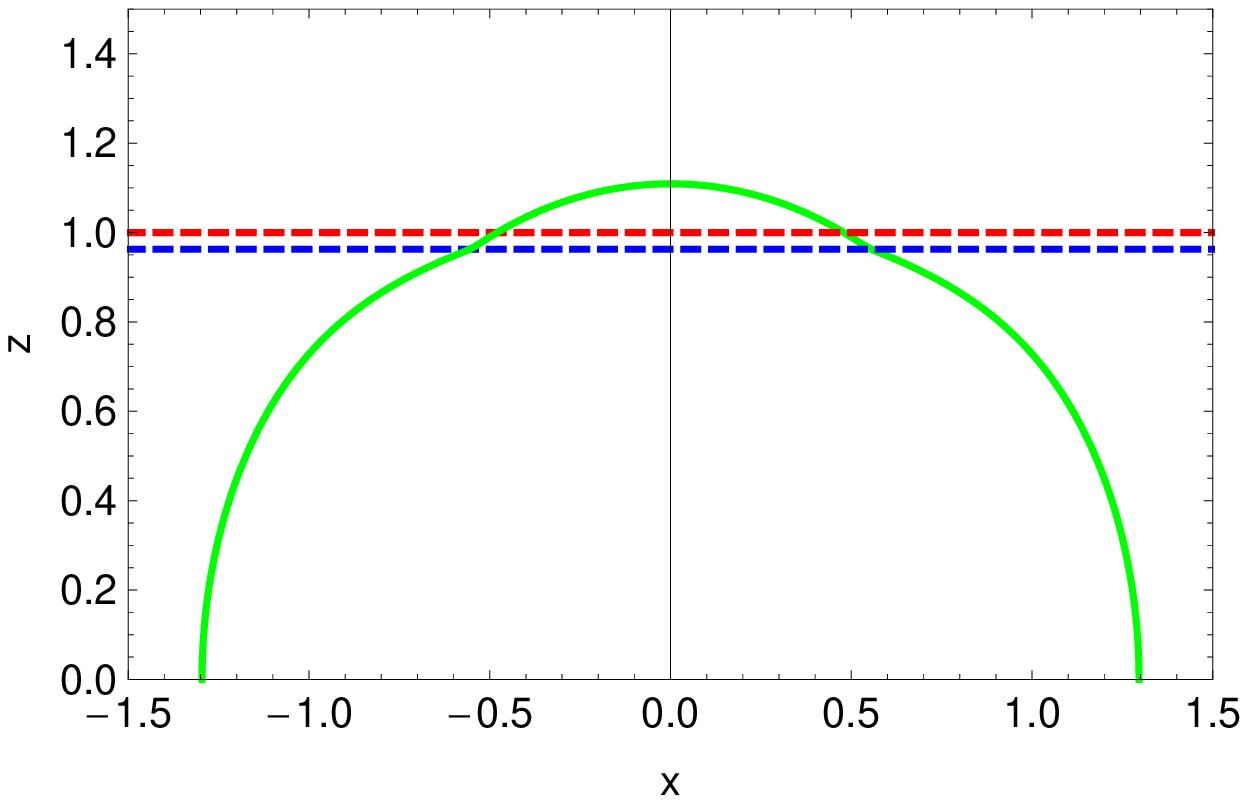}
\label{shell6_RN}
}
\subfigure[$t_0=2.004$ - AdS-Vaidya]{
\includegraphics[scale=0.45]{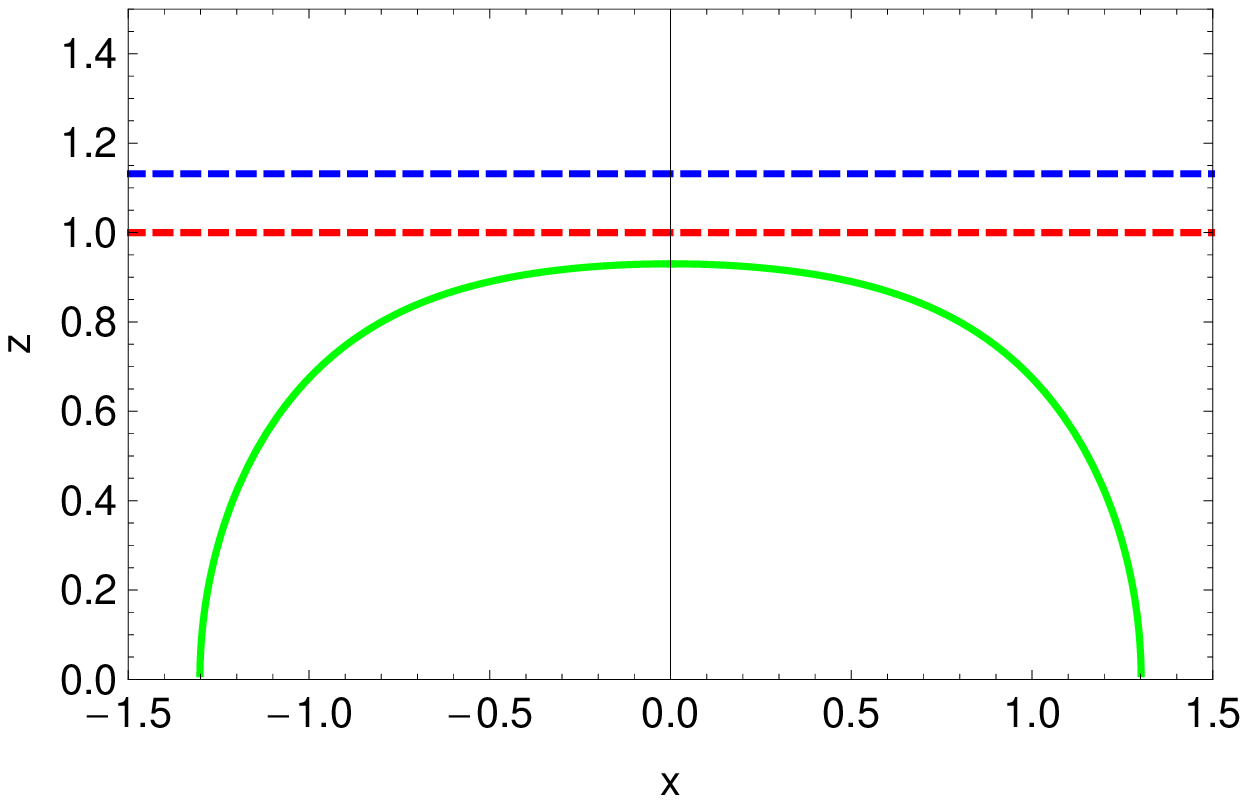}
\label{shell7_RN} } \subfigure[$t_0=2.022$ - RNAdS Vaidya]{
\includegraphics[scale=0.45]{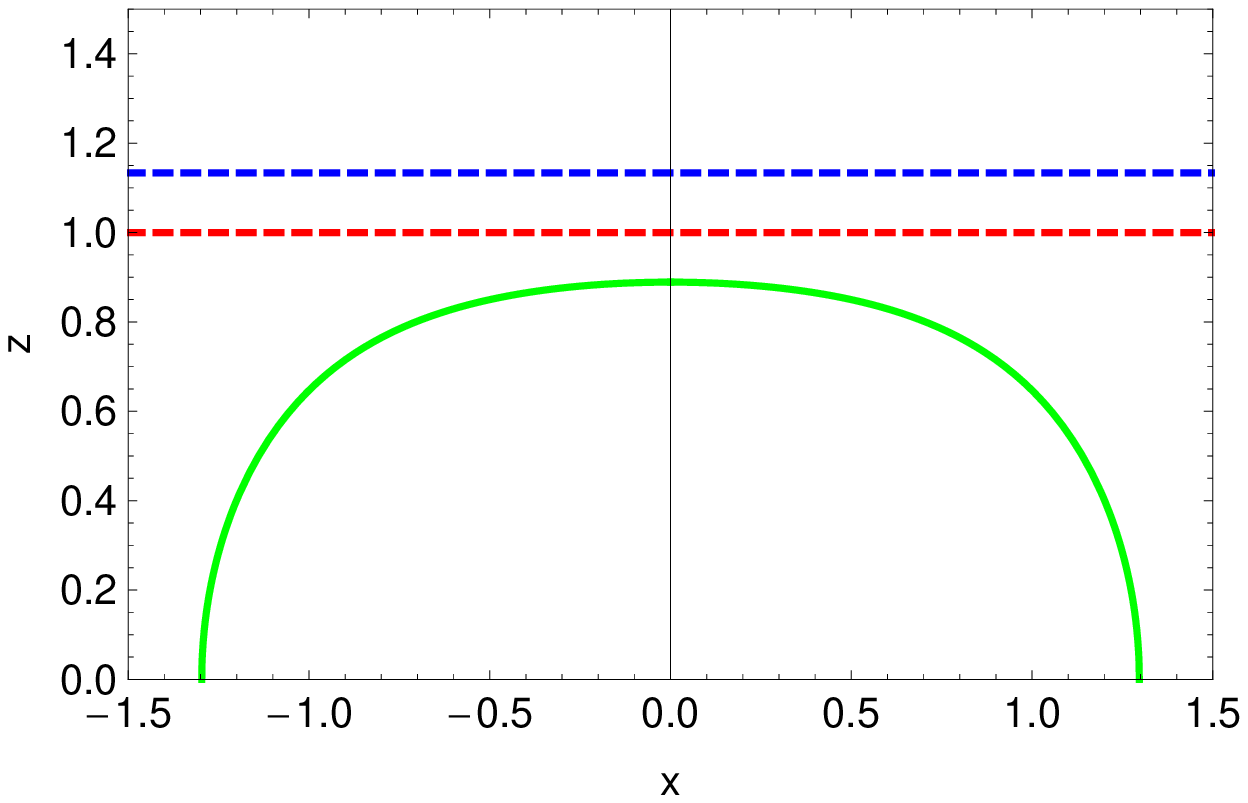}
\label{shell8_RN} } \caption{\small Evolution of the shell of dust
described by the AdS-Vaidya metric with $r_h=1$, compared with a
shell of charged dust described by the RNAdS Vaidya type metric also
with $r_h=1$ and $Q/M=1/2$. In both cases the separation of the
boundary field theory operator pair is $\ell=2.6$. The black hole
horizon is indicated by a dashed horizontal red line. The horizonal
dashed blue lines indicate the position of the shell in each case.
The time differences in each pair of figures is due to the fact that
they have been obtained numerically, and therefore, there are minor
differences.} \label{fig shell advance RN}
\end{figure}

In order to study the thermalization we first have to see the
evolution of the geodesics as shown in Fig. \ref{fig shell advance
RN}. Thus, we solve the problem for a metric with an event horizon
located at $r_h=1/z_h=1$. The difference in comparison with the
AdS-BH case is that now we have to set a non-vanishing charge/mass
ratio. Let us consider $Q/M=1/2$. This choice satisfies the
restrictions from Eqs.(\ref{masa RN}) and (\ref{caso extremo}). From
Fig. \ref{fig shell advance RN} we can see and compare the evolution
of geodesics with both a thin shell of charged and uncharged matter
for $d=4$. In both systems there is a similar behavior. For short
times the geodesics do not change with respect to the pure AdS
space. On the other hand, starting from $t_0 \sim 0.8$, in this
configuration with operator boundary separation $\ell=2.6$,
differences between the charged and uncharged systems become
apparent. Notice that within the range  $0.8<t_0\leq 1$ geodesics in
both systems change abruptly before both systems reach the
thermodynamical equilibrium. Although this process is the same for
the RNAdS-BH case, the actual evolution of the probes is different.
Indeed, we notice differences which render the charged system to
have a slightly larger thermalization time in comparison with the
system at zero chemical potential.

~

~

\subsection{Minimal area surfaces}

We now focus on the thermalization of expectation values of
rectangular Wilson loops. Proceeding in a similar way as for the
geodesic lengths, now we consider the computation of the minimal
area surfaces as described before. Using the AdS-Vaidya type metric,
the Nambu-Goto action becomes,
\begin{eqnarray}
{\cal{A}}_{NG}(t_0,\ell,R)= \frac{R}{2\pi} \int^{\ell/2}_{-\ell/2}
dx \frac{\sqrt{1-f_{RN} (v,z) v'^2-2z'v'}}{z^2} \, .
\label{area_rec_wil_loop}
\end{eqnarray}

We are considering boundary rectangles parametrized by the
coordinates $(x_1,x_2)$. The rest of the coordinates (if there is
any) at the AdS boundary are kept fixed. One assumes the
translational invariance along $x_2$, thus, we will use $x_1$ to
parametrize the functions $v(x_1)$ and $z(x_1)$ in the AdS$_{d+1}$,
and we call it $x$. Along the $x_2$ direction the rectangular path
on the boundary has length $R_{WL}$.

As in the previous case, there is no explicit dependence on $x$ and
therefore, there is a conserved quantity corresponding to the
Hamiltonian. The tip of the rectangular Wilson loop is $z_*$, with
$z'(0) = v'(0) = 0$. Then, the conservation equation becomes
\begin{eqnarray}
1 - 2 z' v' - f_{RN}(v,z) v'^2 = \left( \frac{z_*}{z} \right)^4.
\label{ec_cons_rwl}
\end{eqnarray}

The boundary conditions continue to be the same as in the geodesics case,
\begin{equation}
z(-\ell/2)  =  z_0 \, , \,\,\,\,\,\,\,\, z(\ell/2)   =  z_0 \, ,
\,\,\,\,\,\,\,\, v(-\ell/2)  =  t_0 \, , \,\,\,\,\,\,\,\, v(\ell/2)
=  t_0 \, .
\end{equation}

Next, we have to minimize the Nambu-Goto action for this geometry.
Since the calculations are similar to those for the two-point
functions we will just focus on the final resulting expressions. The
simplified equations of motion of $z(x)$ and $v(x)$ are respectively
\begin{eqnarray}
0 & = & z(x) v'(x)^2 \partial_z f_{RN}(v,z)-4 v'(x)^2
f_{RN}(v,z)-2 z(x) v''(x)-8 v'(x) z'(x) + 4 \, , \label{eom1wl} \\
0 &=& v'(x) z'(x) \partial_z f_{RN}(v,z) + \frac{1}{2} v'(x)^2
\partial_v f_{RN}(v,z) + v''(x) f_{RN}(v,z) + z''(x) \, .
\label{eom2wl}
\end{eqnarray}
Using the explicit form of $f_{RN}(v,z)$ we obtain the following set
of differential equations
\begin{eqnarray}
0 & = & 4 z(x)^d m(v) v'(x)^2-4 z(x)^{2 d-2} q(v)^2 v'(x)^2-2 z(x) v''(x) \nonumber \\
&&  -8 v'(x) z'(x)-4 v'(x)^2+4 \, , \label{eom1wlRN} \\
0 &=& \frac{1}{2} z(x)^d v'(x)^2 m'(v)+z(x)^d m(v) v''(x)+d
z(x)^{d-1} m(v) v'(x) z'(x) \nonumber \\
&& -z(x)^{2 d-2} q(v) v'(x)^2 q'(v)-z(x)^{2 d-2} q(v)^2 v''(x)-2 d
z(x)^{2 d-3} q(v)^2 v'(x) z'(x) \nonumber \\
&& +  2 z(x)^{2 d-3} q(v)^2 v'(x) z'(x)-v''(x)-z''(x) \, ,
\label{eom2wlRN}
\end{eqnarray}
for $z(x)$ and $v(x)$, respectively, which we will solved
numerically. These equations reduce to the corresponding ones
presented in \cite{Balasubramanian:2011ur} when the charge is set to
zero.

We can again extract the information of time and boundary separation
length from
\begin{eqnarray}
z(\ell/2)=z_0 \, ,  \hspace{1.5cm} v(\ell/2)=t_0 \, ,
\end{eqnarray}
and rewrite the on-shell Nambu-Goto action by making use of the
conservation equation, obtaining
\begin{eqnarray}
{\cal{A}}(t_0,\ell,R_{WL})= \frac{R_{WL}}{\pi} \int_{0}^{\ell/2} dx
\, \frac{z_*^2}{z^4} \, .
\end{eqnarray}
Finally, we subtract the divergent part from pure AdS space by
defining
\begin{eqnarray}
\delta {\cal{A}}(t_0,\ell,R_{WL}) = {\cal{A}}(t_0,\ell,R_{WL}) -
\frac{1}{z_0} \frac{R_{WL}}{\pi} \, .
\end{eqnarray}
%

\subsection{Evolution of thermalization probes}

Having the information about the geodesic lengths and minimal area
surfaces as a function of time, we can describe the thermalization
process. In order to do it we compare $\delta {\cal{L}}$ and $\delta
{\cal{A}}$ at each time with the final values $\delta
{\cal{L}}_{RN}$ and $\delta {\cal{A}}_{RN}$, obtained in the pure
Reissner-Nordstr\"om-AdS black hole geometry, {\it i.e.} by setting
both $m(v)=M$ and $q(v)=Q$ constant. This procedure is repeated for
different AdS-boundary separations $\ell$ and different spacetime
dimensions. We prefer to plot the quantities $\bar{\delta {\cal{L}}}
\equiv \delta {\cal{L}}/\ell$ and $\delta \bar{{\cal{A}}} \equiv
\delta {\cal{A}}/ (R_{WL}\ell/\pi)$, such that the thermalization
effect can be better observed, independently of $\ell$.

Let us consider a systematic study on how the thermalization time
changes within these geometries. If we want to have a clear
comparison with thermalization in the Schwarzschild-AdS case without
charge, it is natural to keep fixed the radius of the event horizon,
say $r_h=1$. The mass of the black hole is given by Eq.(\ref{masa
RN}), so we obtain
\begin{eqnarray}
M = 1 + Q^2 \, . \label{ec masa r unidad}
\end{eqnarray}
This relation between $M$ and $Q$ holds from $Q=0$ up to the
extremal case. From Eq.(\ref{caso extremo}), we find
\begin{eqnarray}
Q_{ext} = \sqrt{\frac{d}{d-2}} \, .
\end{eqnarray}
Eq.(\ref{ec masa r unidad}) still holds for $Q$ values larger than
its extremal value. However, in that case $r_h=1$ becomes a hidden
singularity. The black hole horizon is the larger squared root of
the function $U(r)$ as given by Eq.(\ref{u de erre}). With this
choice we can explore the whole parameter range of the problem. The
idea is to keep fixed the event horizon radius, and consider the
charge in the range $0 \leq Q \leq Q_{ext}$. Then, we will study all
the range $0\leq \mu/T \leq \infty$ in the dual field theory.

This is shown for the two-point functions for different spacetime
dimensions $d=3, 4, 5, 6$ with $\ell=2, 3, 4$ in Fig. \ref{fig
termalizacion tp RN} and for the expectation value of rectangular
Wilson loops for $\ell=1,2,3$ in Fig. \ref{fig wl RN}. Firstly, we
notice certain similarities in comparison with the corresponding
Schwarzschild-AdS cases ($Q=0$). In principle, the thermalization
time increases as the separation of the CFT operators at the
boundary is increased. Moreover, at a fixed value of the charge, the
thermalization time increases as the boundary separation does. For
instance, for $\ell=2$, the renormalized geodesic length practically
has no significant differences for the thermalization time for any
charge and dimension. However, interestingly, for $\ell=3$, the
differences become more evident, while for $\ell=4$ RNAdS-BH and
AdS-BH cases are notoriously distinct.

\begin{figure}
\centering \subfigure[AdS$_4$ $\&$ $\ell=2$]{
\includegraphics[scale=0.4]{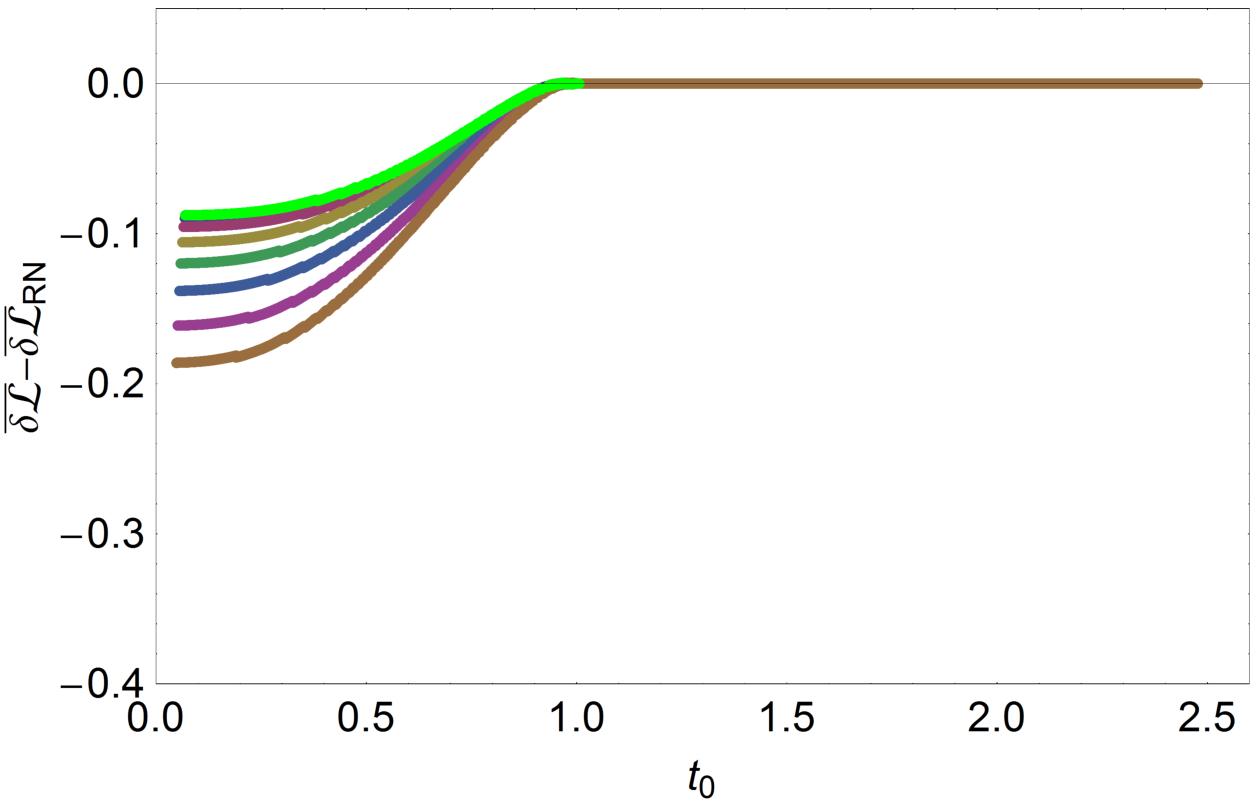}
\label{t2_ads4_RN} } \subfigure[AdS$_4$ $\&$ $\ell=3$]{
\includegraphics[scale=0.4]{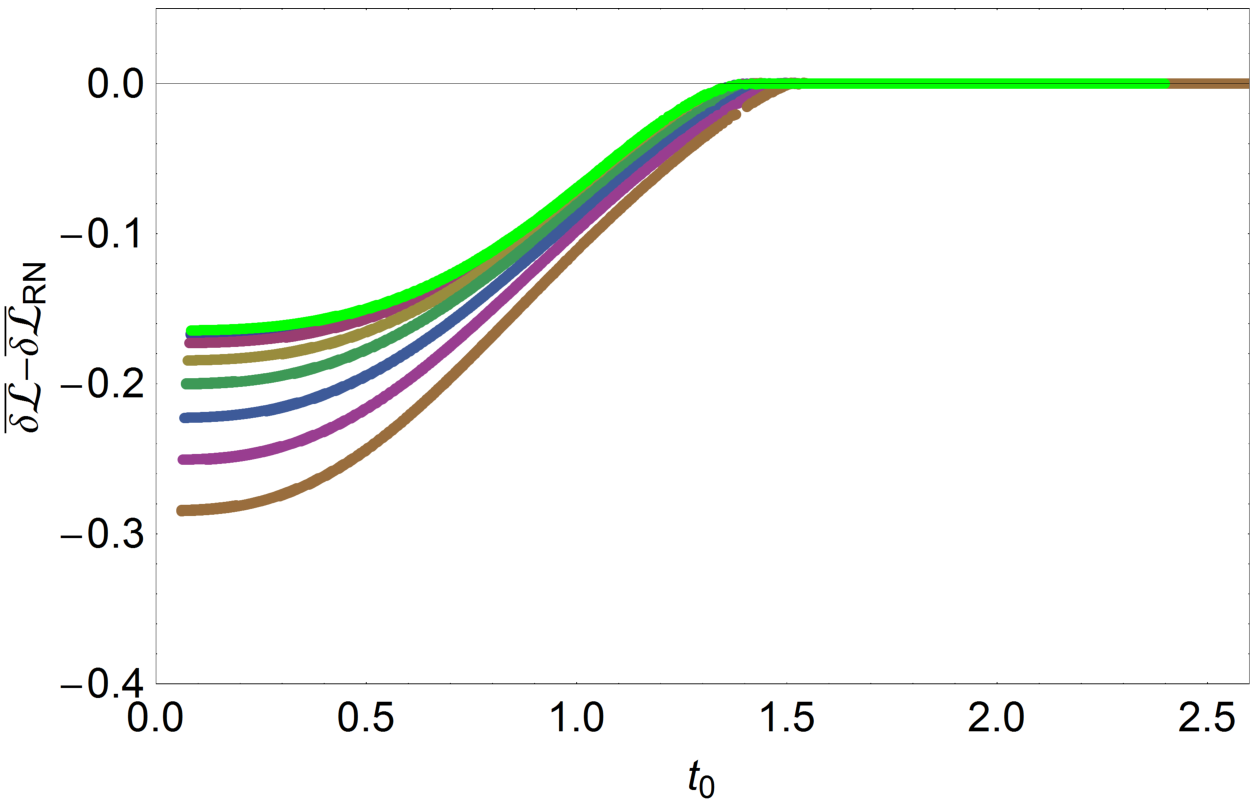}
\label{t3_ads4_RN} } \subfigure[AdS$_4$ $\&$ $\ell=4$]{
\includegraphics[scale=0.4]{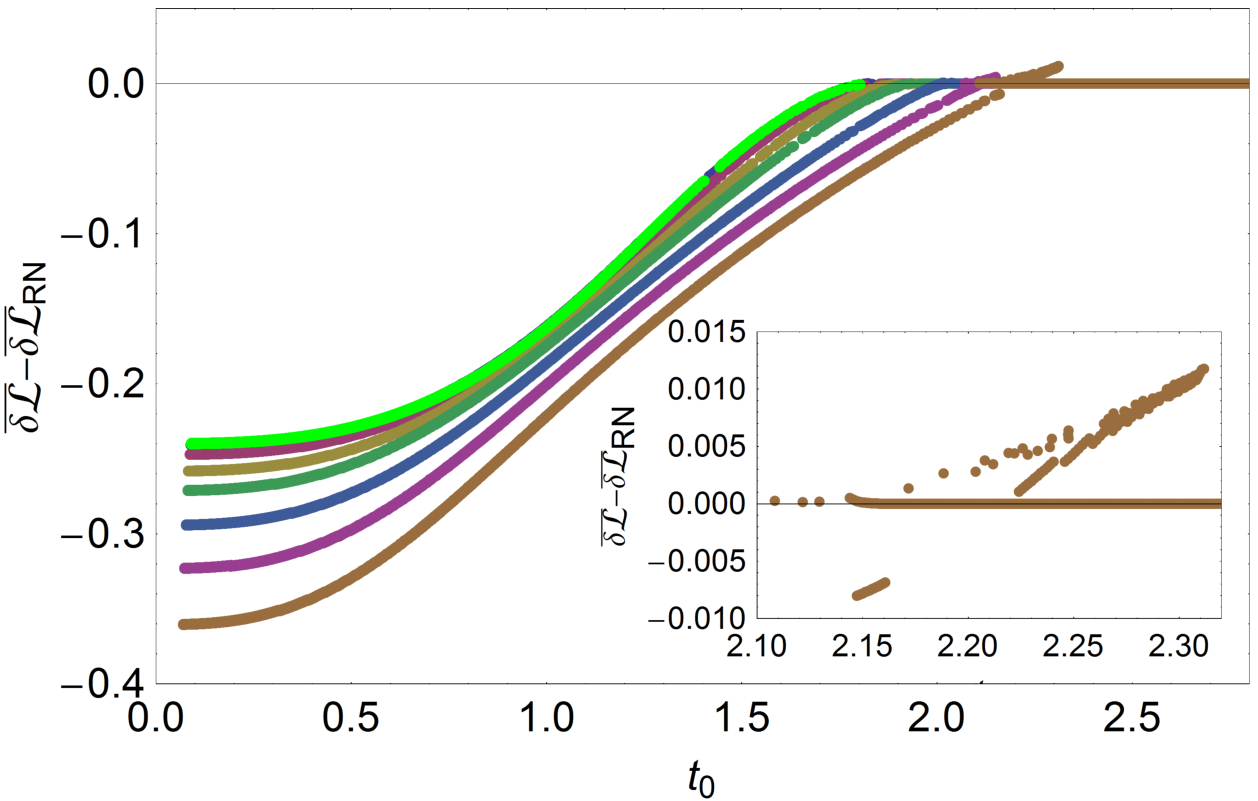}
\label{t4_ads4_RN} } \subfigure[AdS$_5$ $\&$ $\ell=2$]{
\includegraphics[scale=0.4]{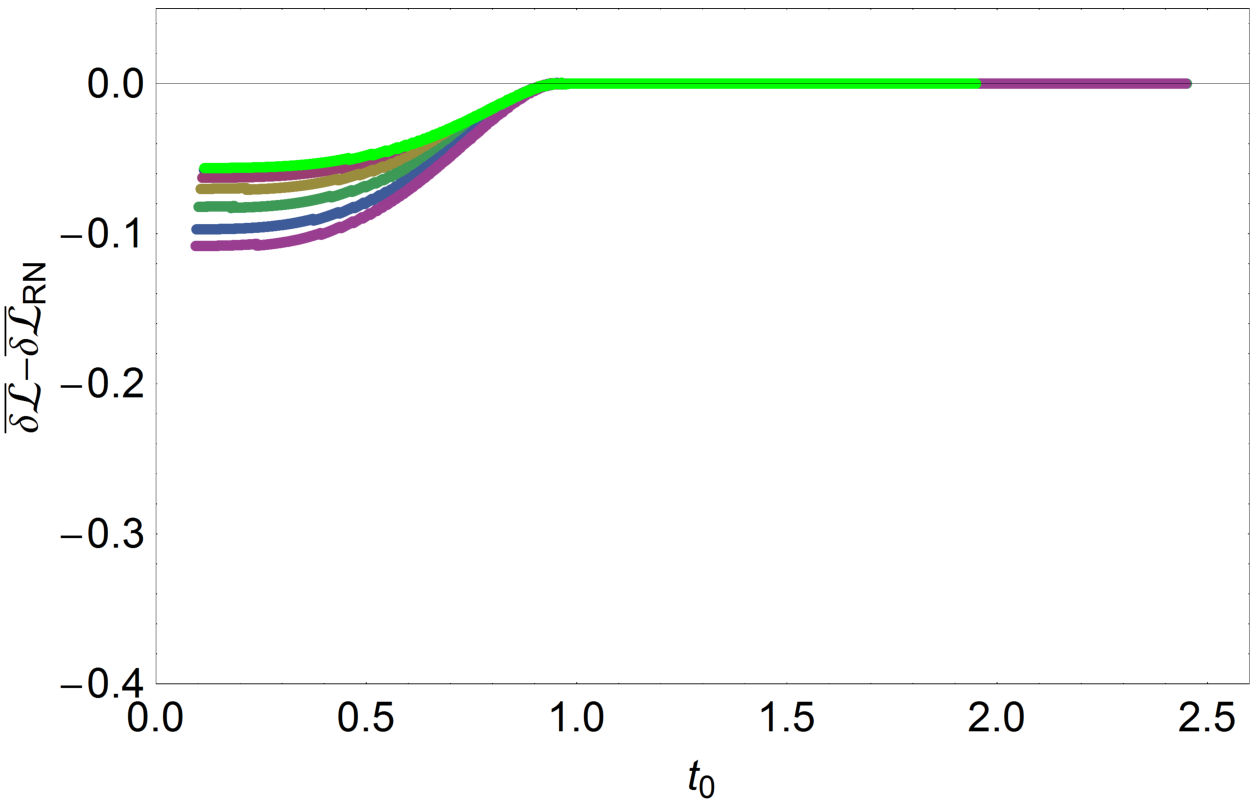}
\label{t2_ads5_RN} } \subfigure[AdS$_5$ $\&$ $\ell=3$]{
\includegraphics[scale=0.4]{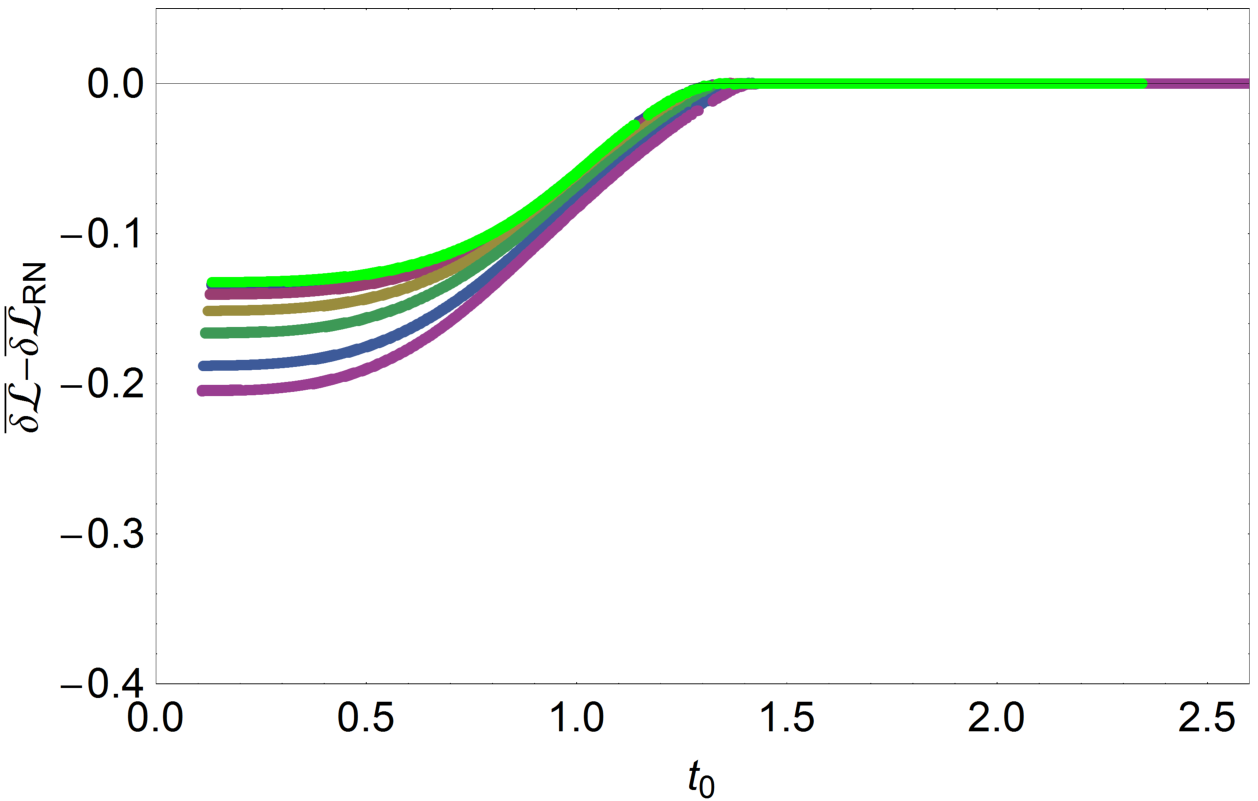}
\label{t3_ads5_RN} } \subfigure[AdS$_5$ $\&$ $\ell=4$]{
\includegraphics[scale=0.4]{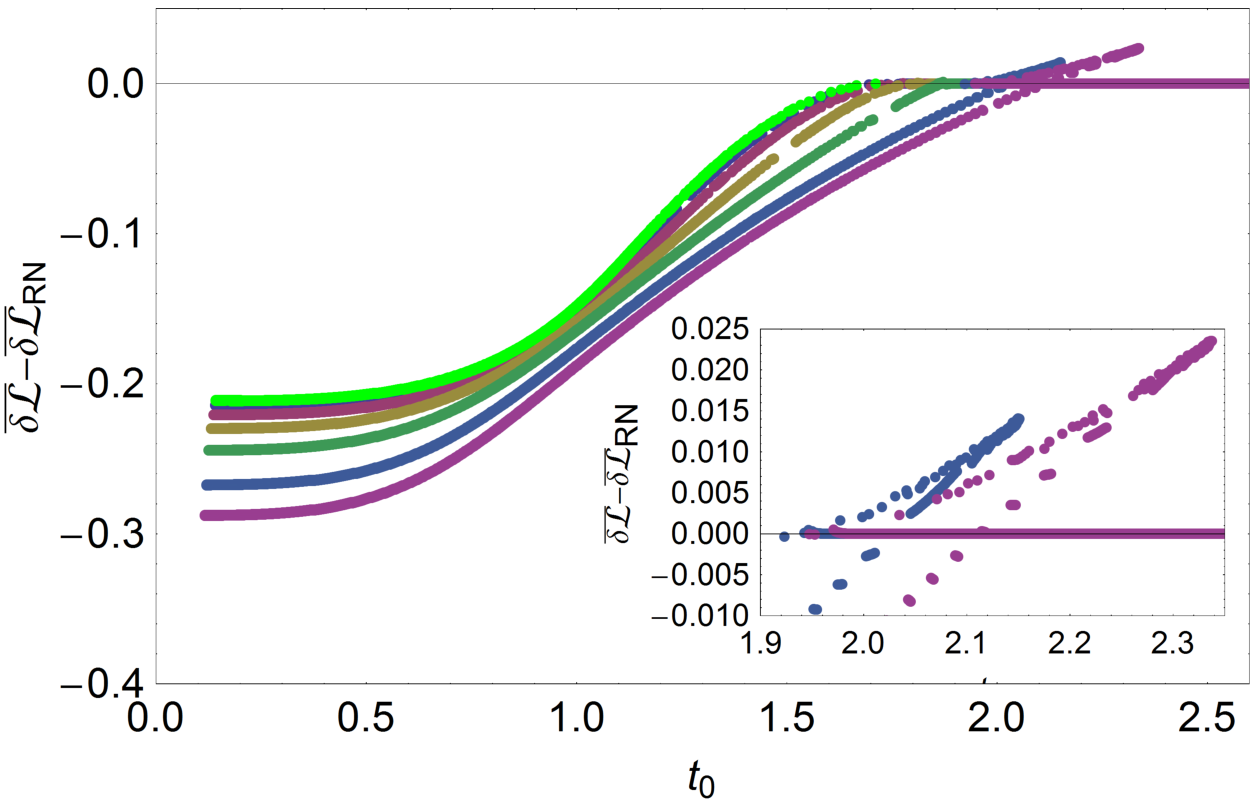}
\label{t4_ads5_RN} } \subfigure[AdS$_6$ $\&$ $\ell=2$]{
\includegraphics[scale=0.4]{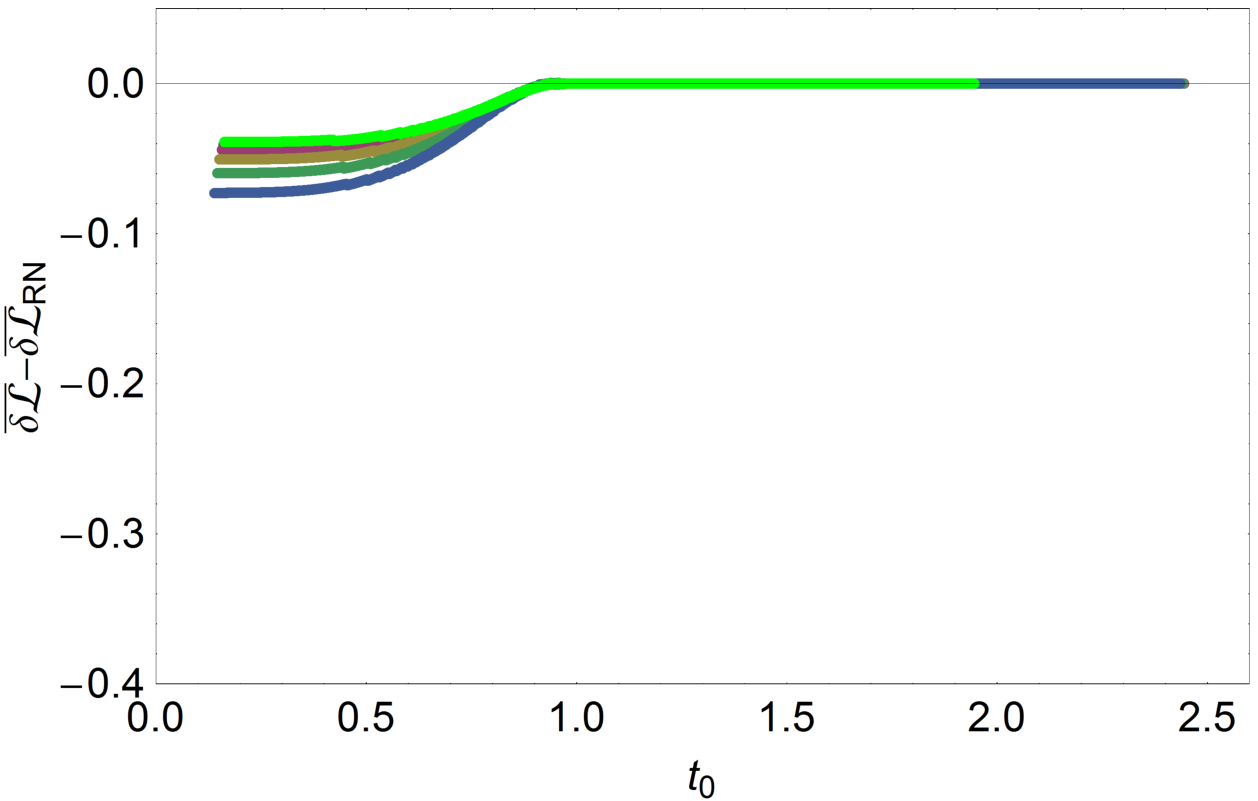}
\label{t2_ads6_RN} } \subfigure[AdS$_6$ $\&$ $\ell=3$]{
\includegraphics[scale=0.4]{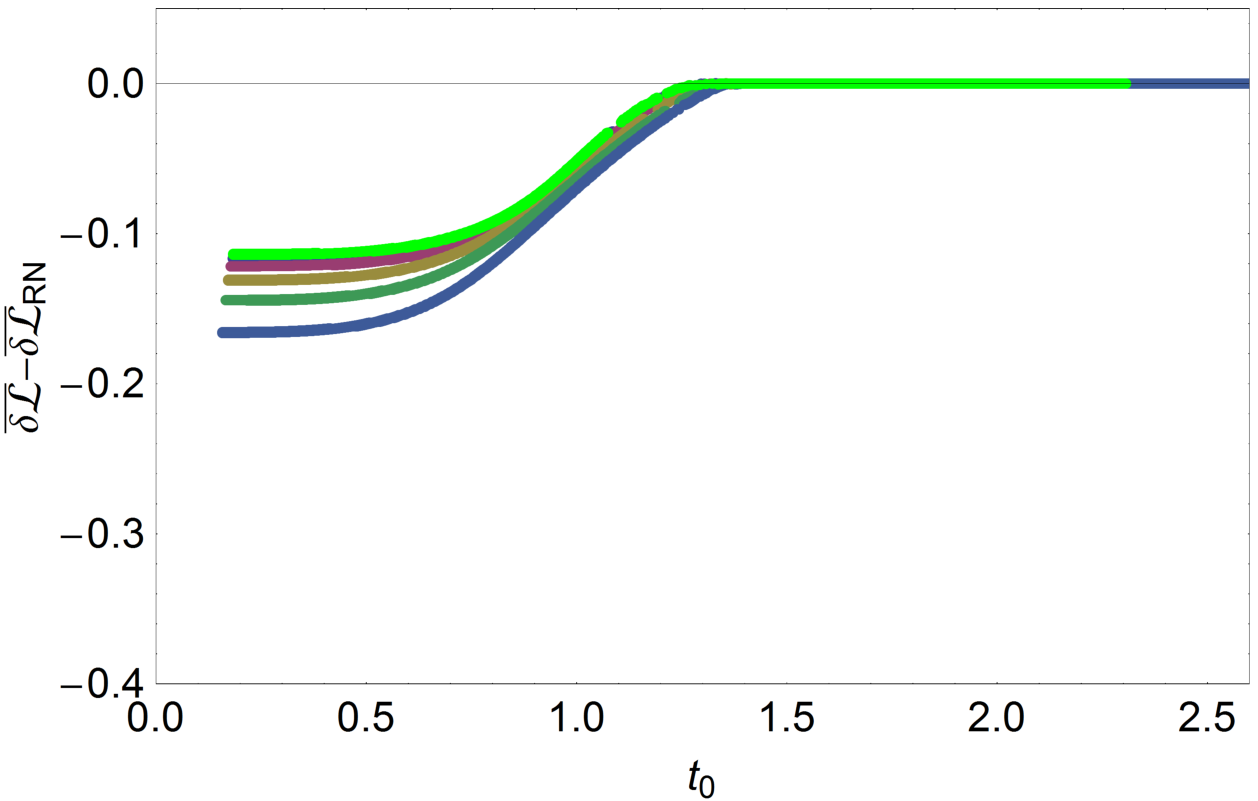}
\label{t3_ads6_RN} } \subfigure[AdS$_6$ $\&$ $\ell=4$]{
\includegraphics[scale=0.4]{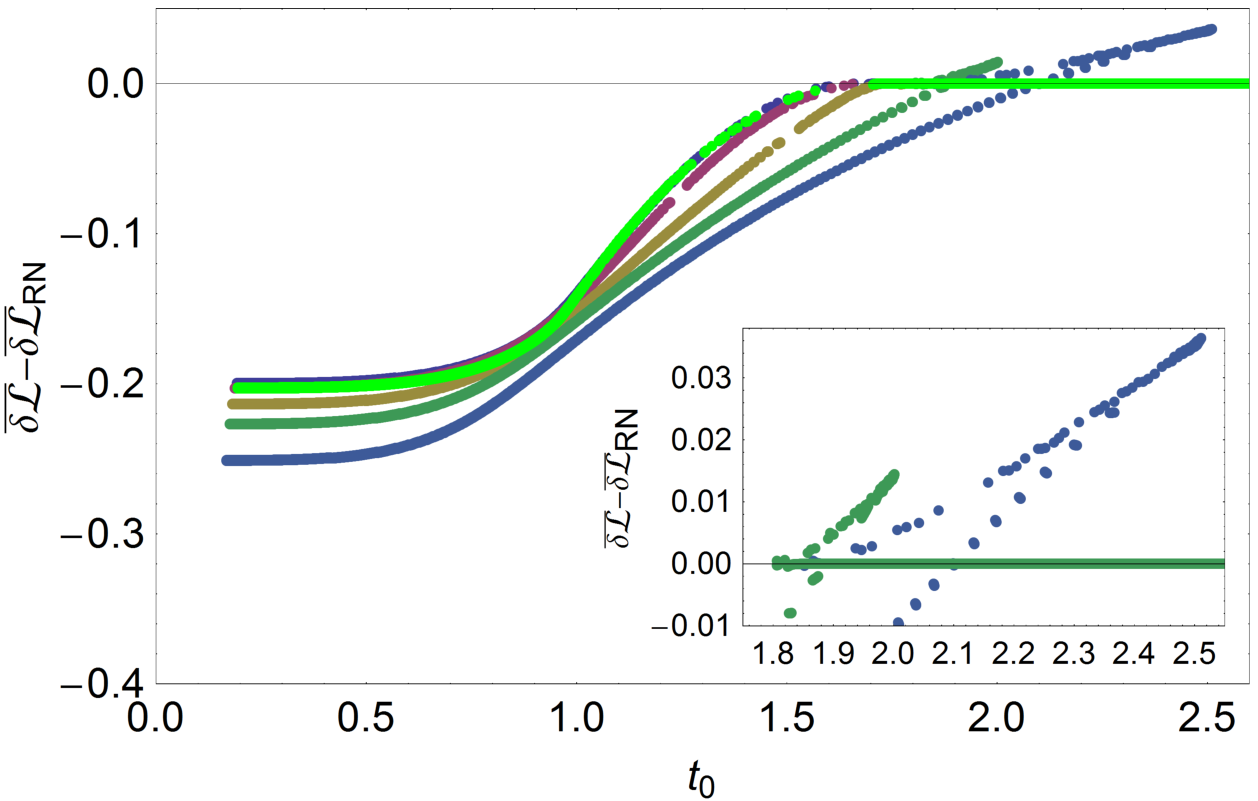}
\label{t4_ads6_RN} } \subfigure[AdS$_7$ $\&$ $\ell=2$]{
\includegraphics[scale=0.4]{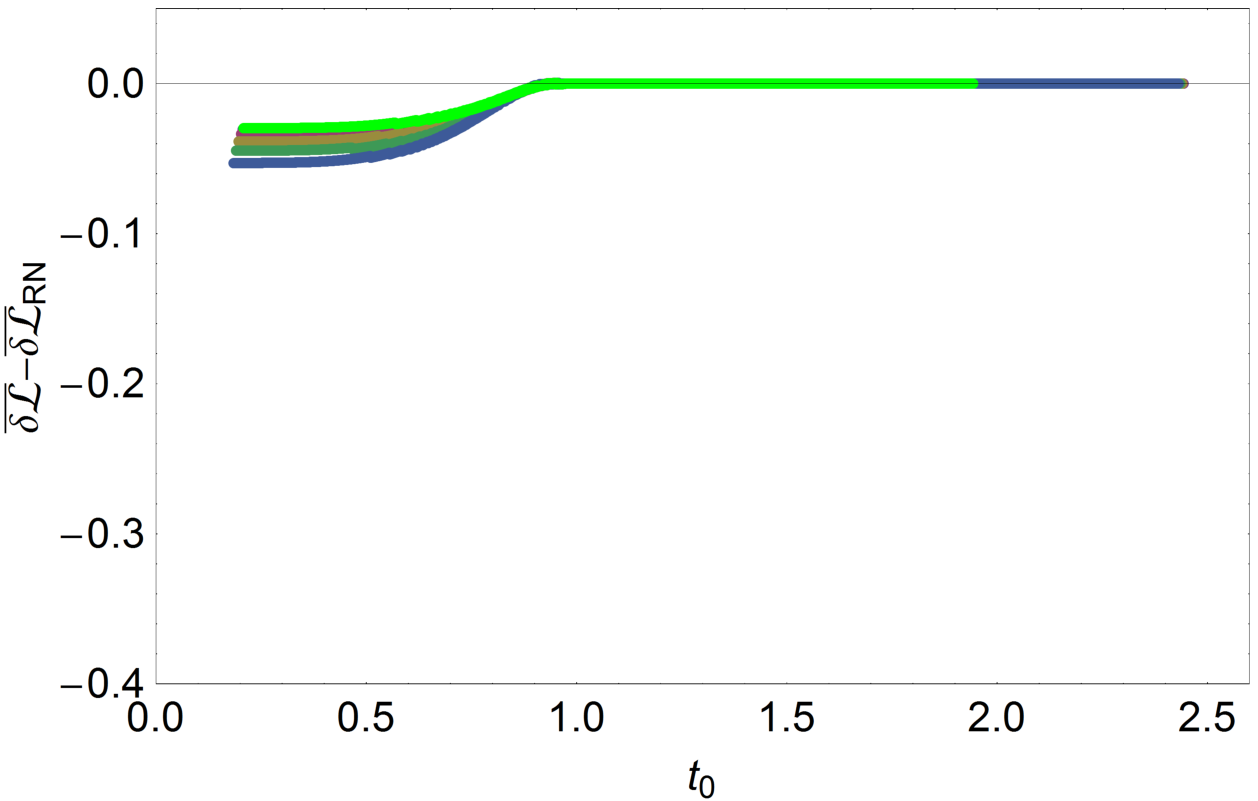}
\label{t2_ads7_RN} } \subfigure[AdS$_7$ $\&$ $\ell=3$]{
\includegraphics[scale=0.4]{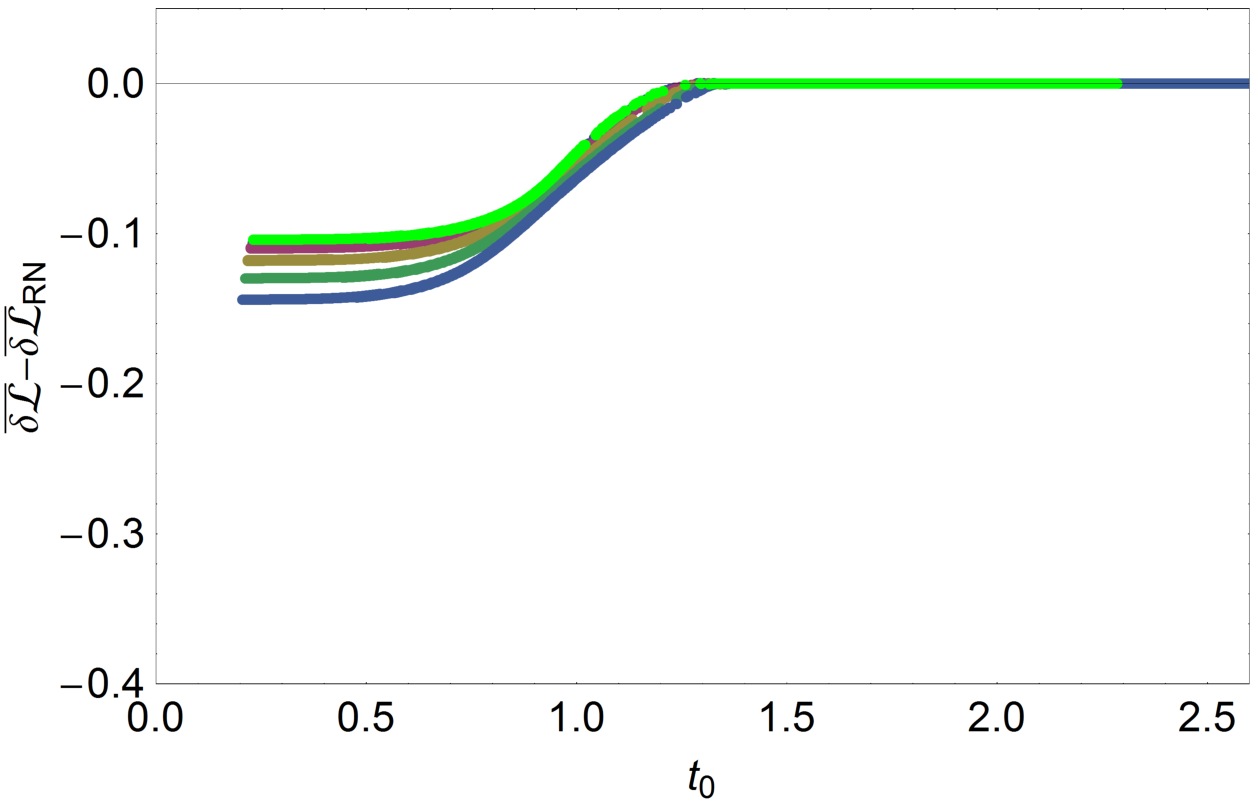}
\label{t3_ads7_RN} } \subfigure[AdS$_7$ $\&$ $\ell=4$]{
\includegraphics[scale=0.4]{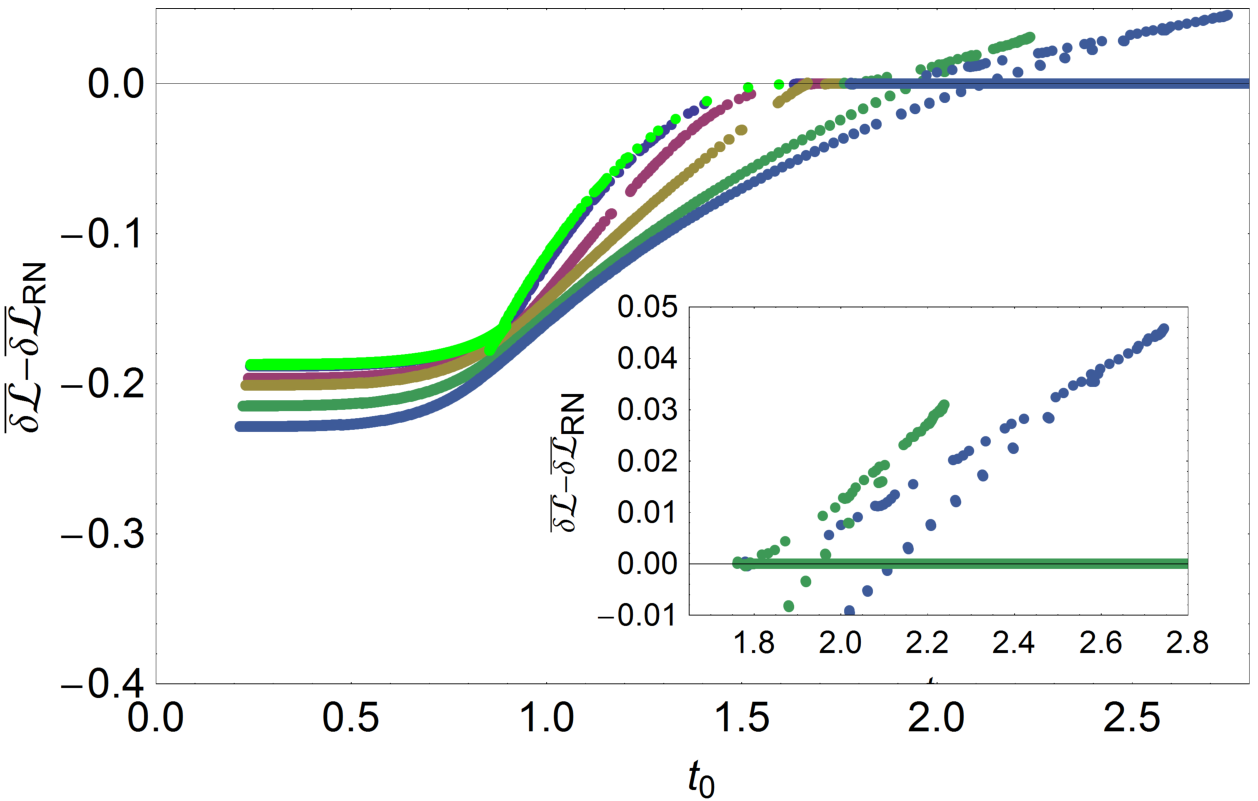}
\label{t4_ads7_RN} } \caption{\small Thermalization of the
renormalized geodesic lengths for a RNAdS Vaidya type metric with
$d+1=4, 5, 6, 7$ and the boundary separation $\ell=2,3,4$. The first
curve in each picture (fluor green) indicates the case $Q=0$,  {\it
i.e.} the Schwarzschild-AdS case. The last curve in each case
corresponds to the extremal case where $\mu/T\rightarrow\infty$. For
$\ell=4$ swallow tails appear before the thermalization takes place.
They are shown with more detail in the corresponding insets.}
\label{fig termalizacion tp RN}
\end{figure}

\begin{figure}
\centering \subfigure[AdS$_4$ $\&$ $\ell=1$]{
\includegraphics[scale=0.4]{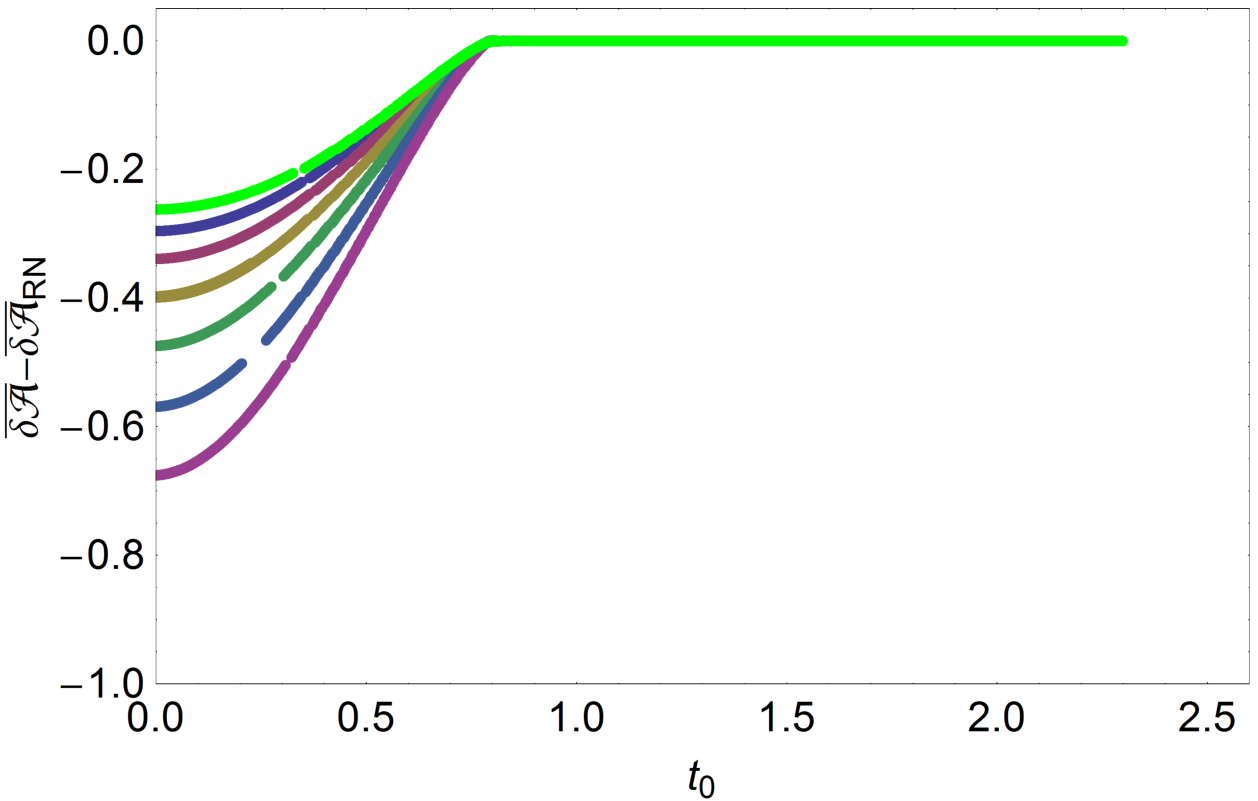}
\label{w1_ads4_RN} } \subfigure[AdS$_4$ $\&$ $\ell=1.5$]{
\includegraphics[scale=0.4]{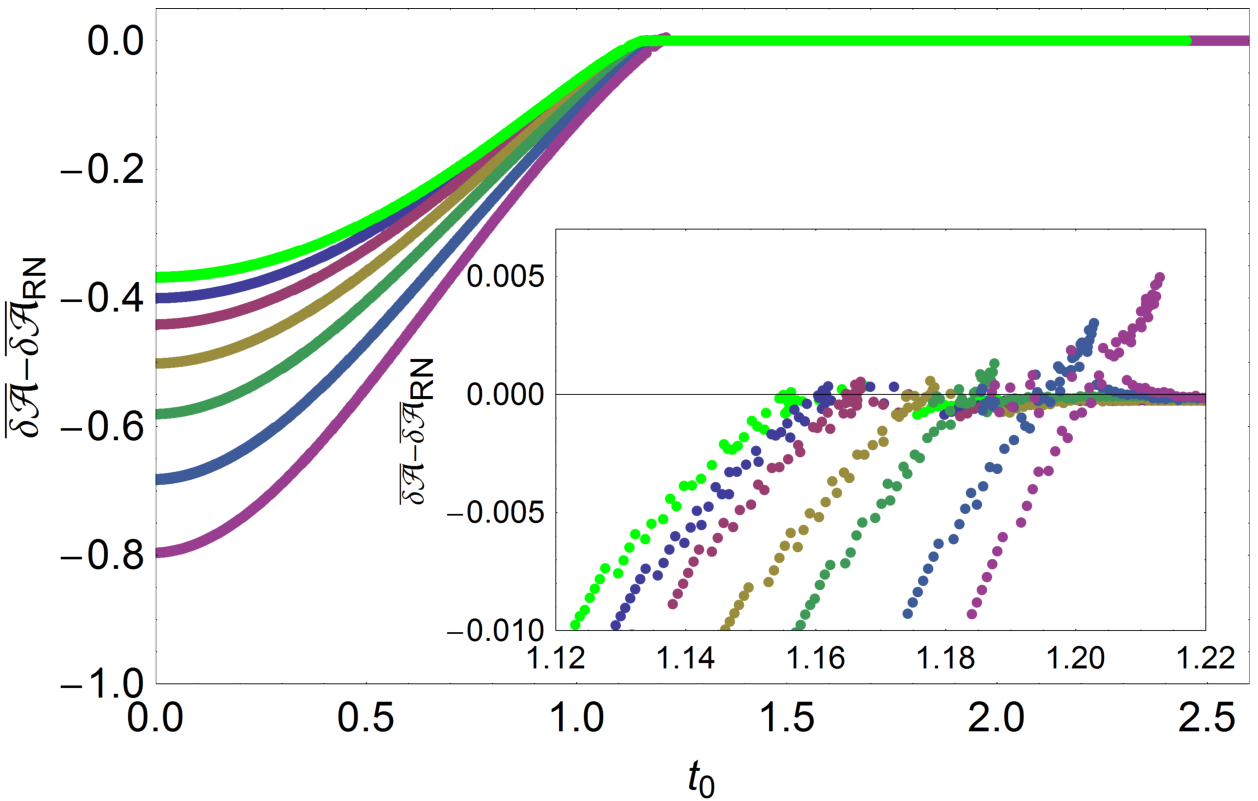}
\label{w15_ads4_RN} } \subfigure[AdS$_4$ $\&$ $\ell=2$]{
\includegraphics[scale=0.4]{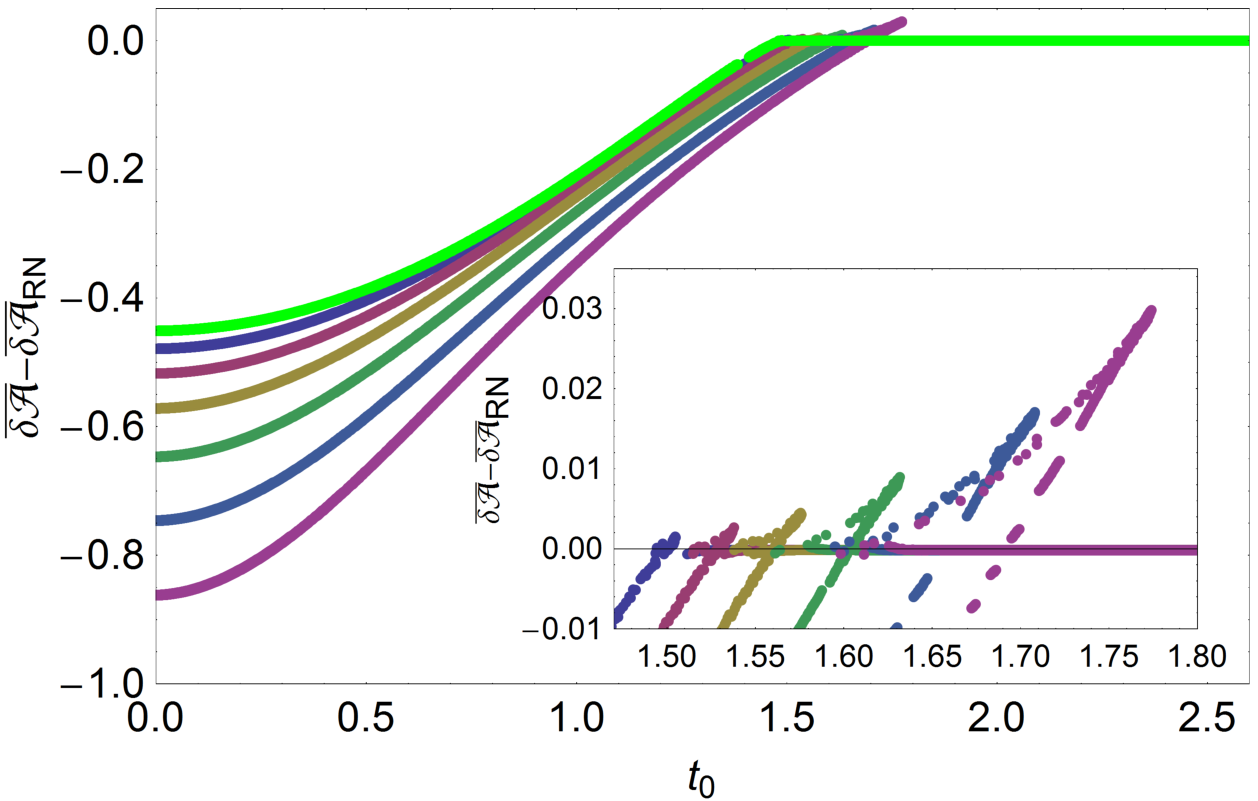}
\label{w2_ads4_RN} } \subfigure[AdS$_5$ $\&$ $\ell=1$]{
\includegraphics[scale=0.4]{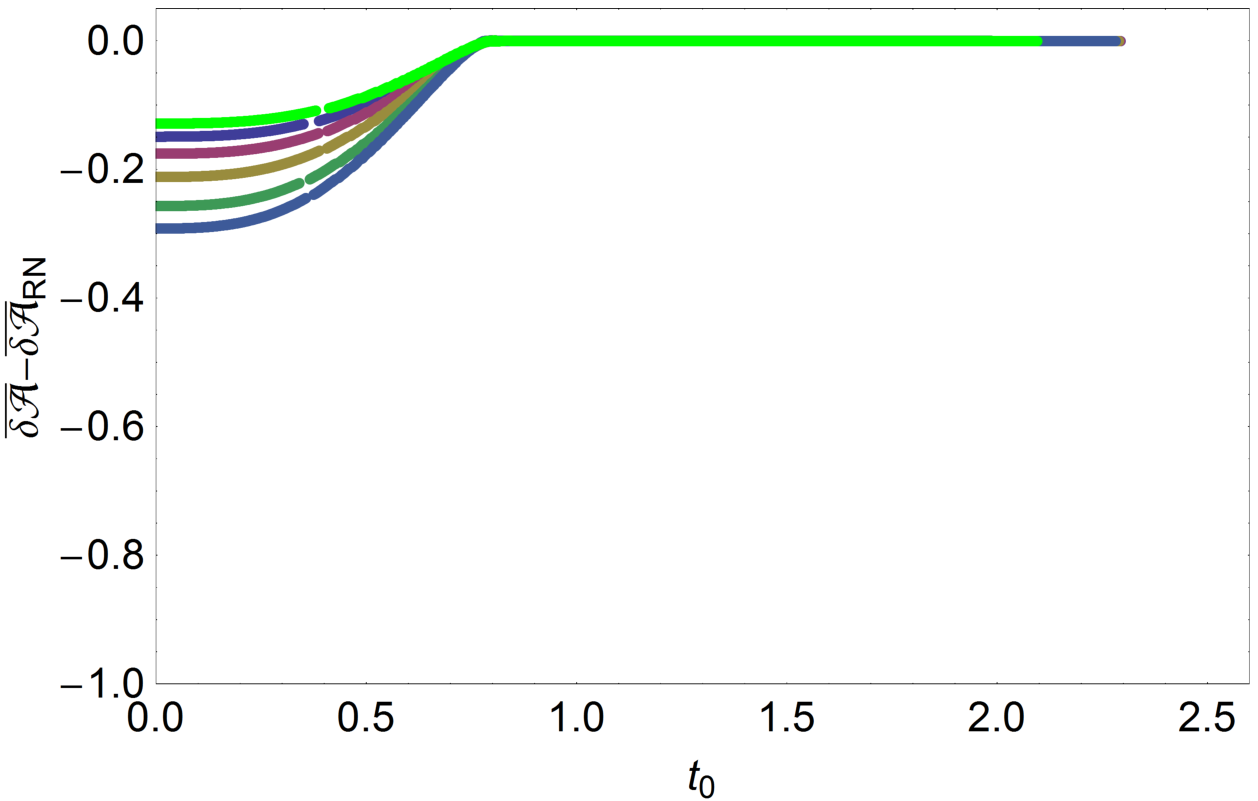}
\label{w1_ads5_RN} } \subfigure[AdS$_5$ $\&$ $\ell=1.5$]{
\includegraphics[scale=0.4]{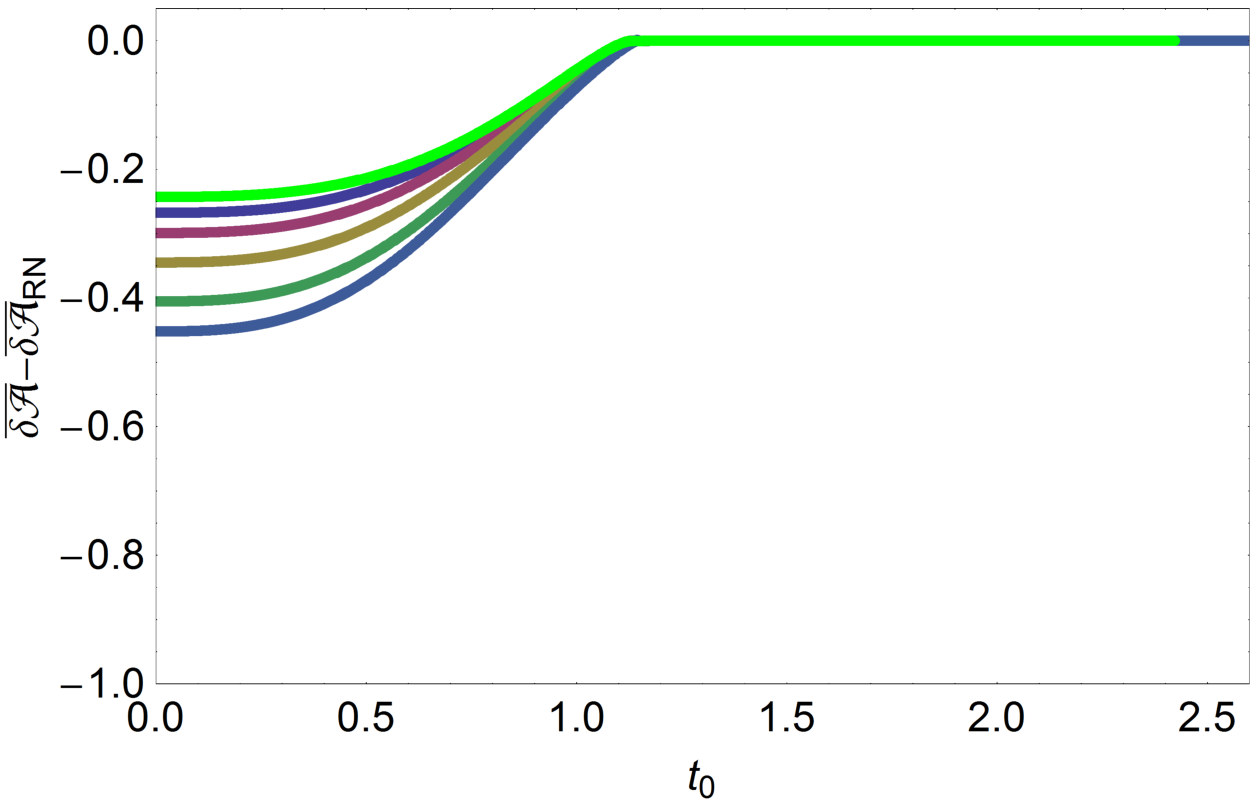}
\label{w15_ads5_RN} } \subfigure[AdS$_5$ $\&$ $\ell=2$]{
\includegraphics[scale=0.4]{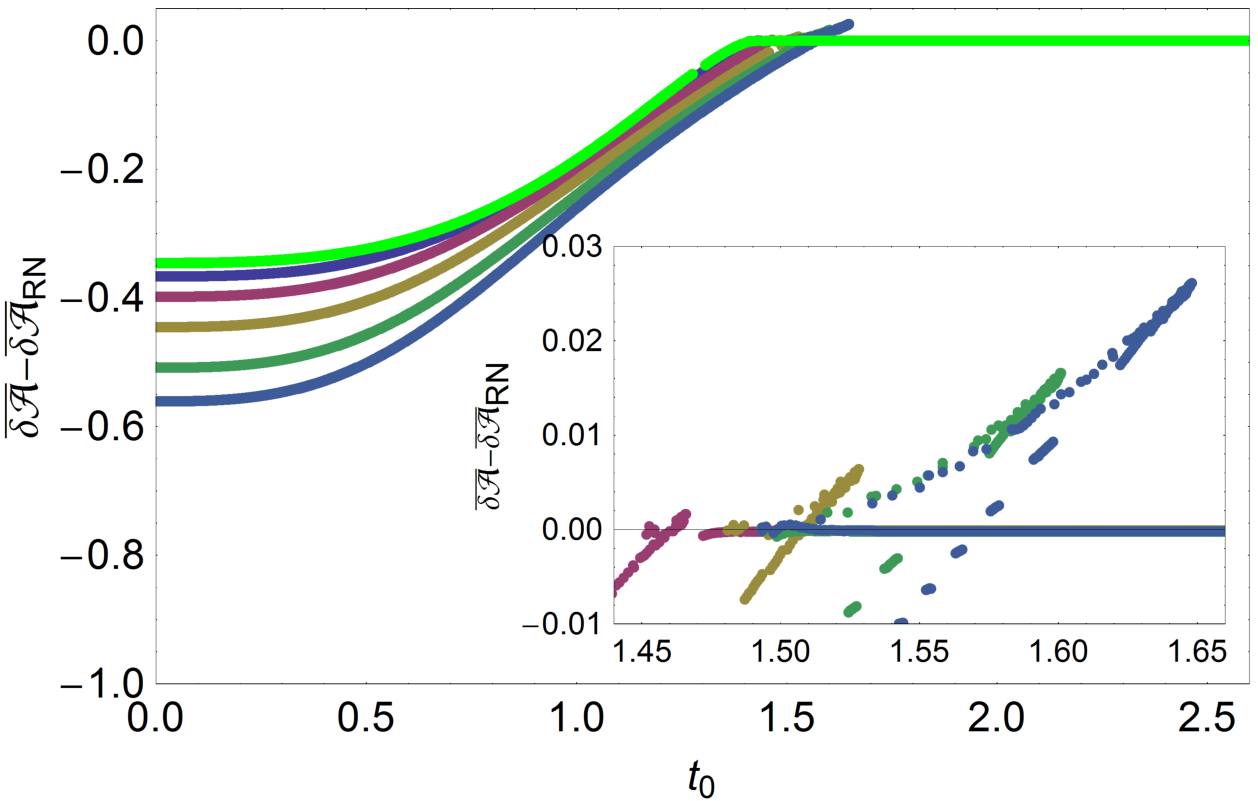}
\label{w2_ads5_RN} } \subfigure[AdS$_6$ $\&$ $\ell=1$]{
\includegraphics[scale=0.4]{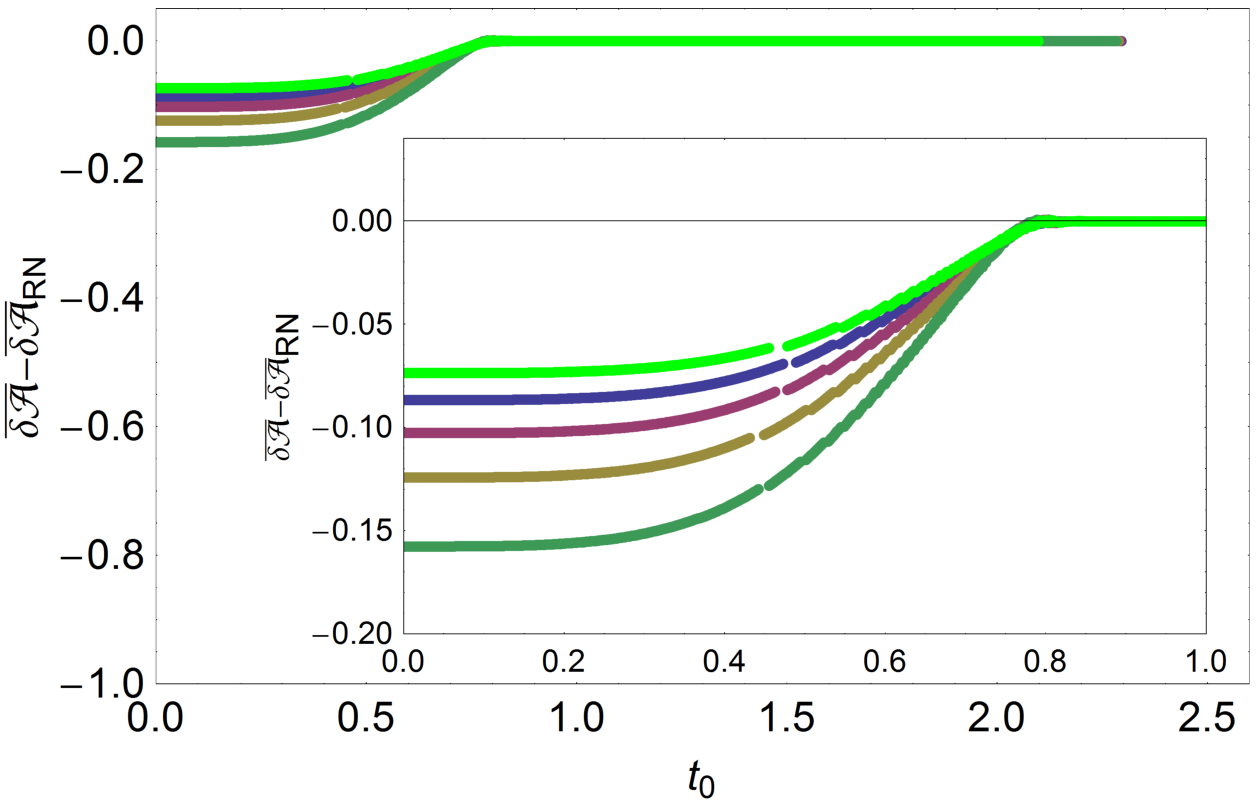}
\label{w1_ads6_RN} } \subfigure[AdS$_6$ $\&$ $\ell=1.5$]{
\includegraphics[scale=0.4]{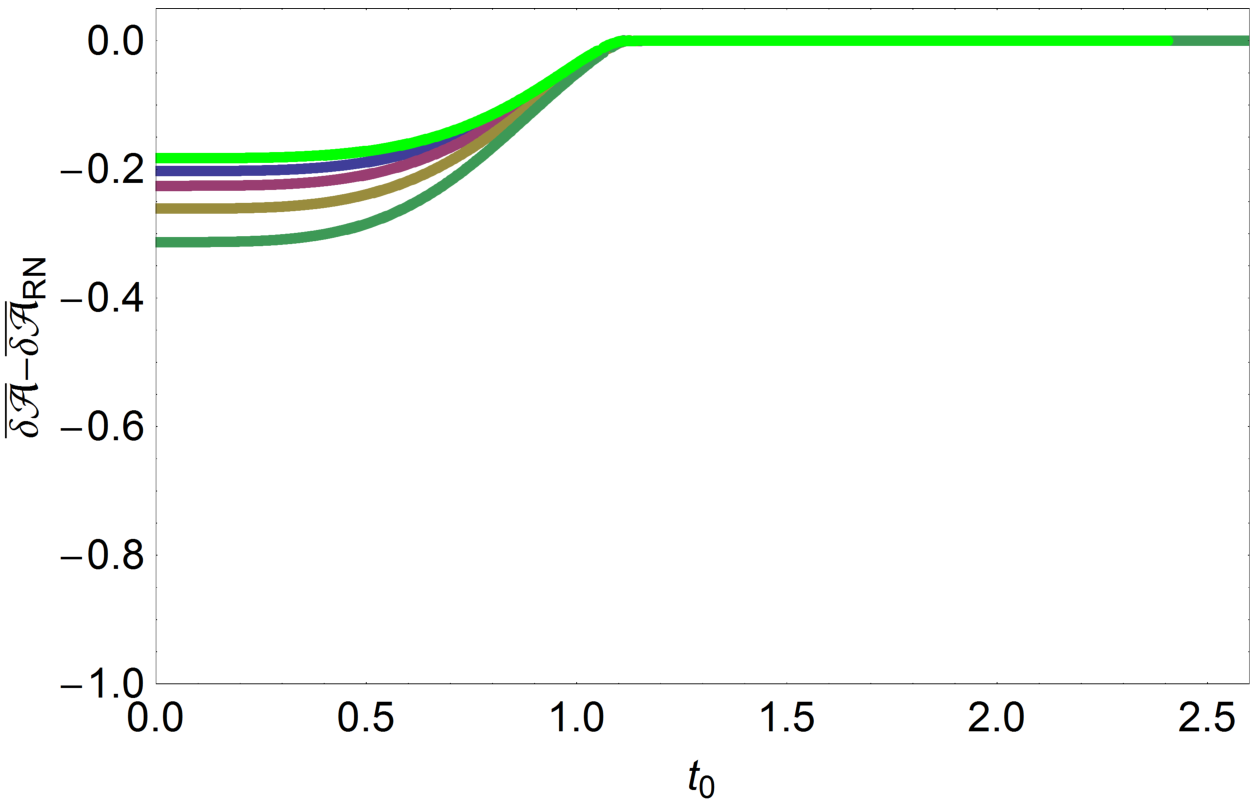}
\label{w15_ads6_RN} } \subfigure[AdS$_6$ $\&$ $\ell=2$]{
\includegraphics[scale=0.4]{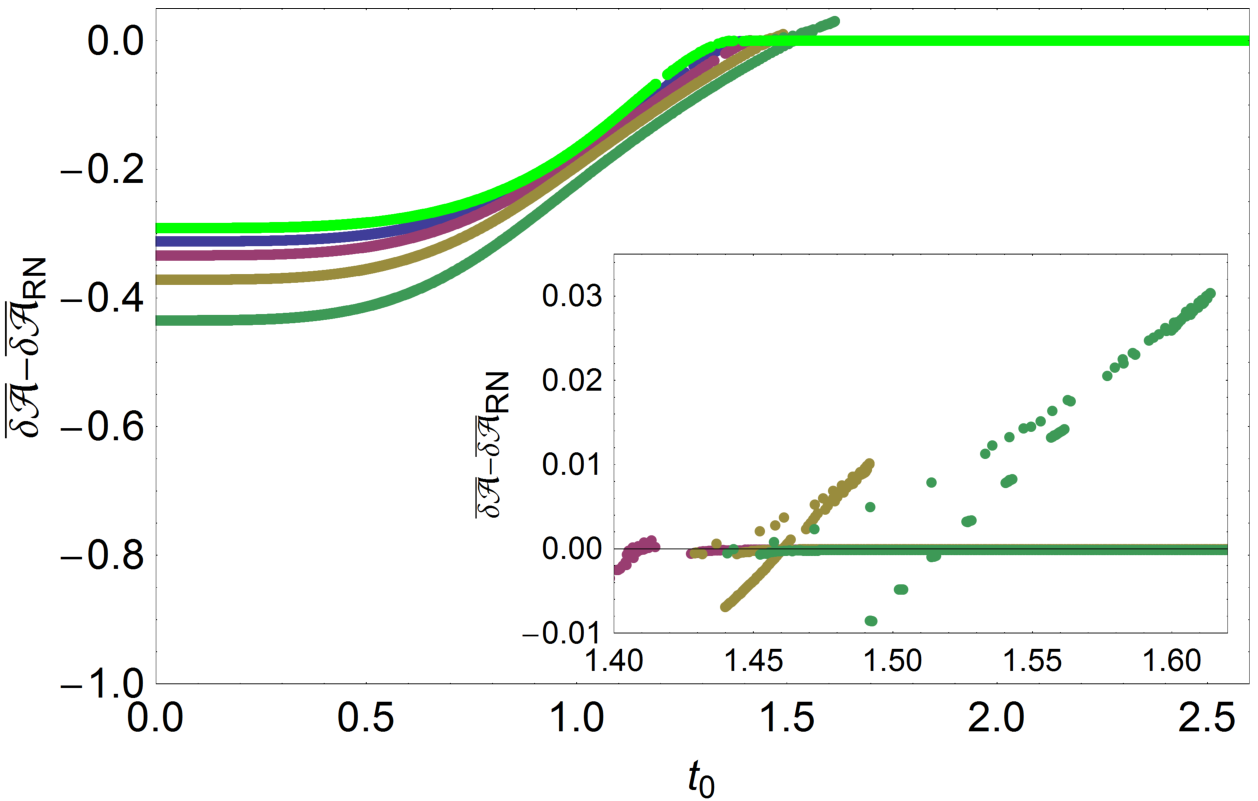}
\label{w2_ads6_RN} } \subfigure[AdS$_7$ $\&$ $\ell=1$]{
\includegraphics[scale=0.4]{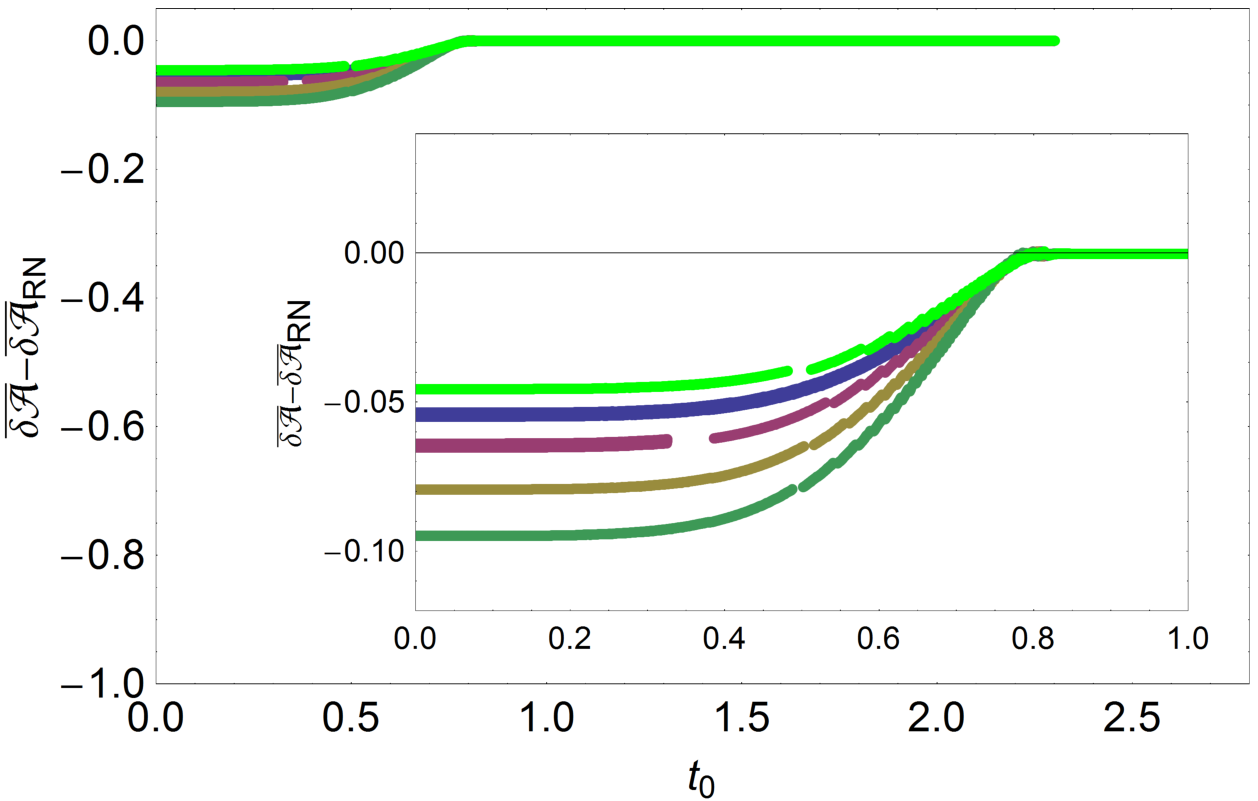}
\label{w1_ads7_RN} } \subfigure[AdS$_7$ $\&$ $\ell=1.5$]{
\includegraphics[scale=0.4]{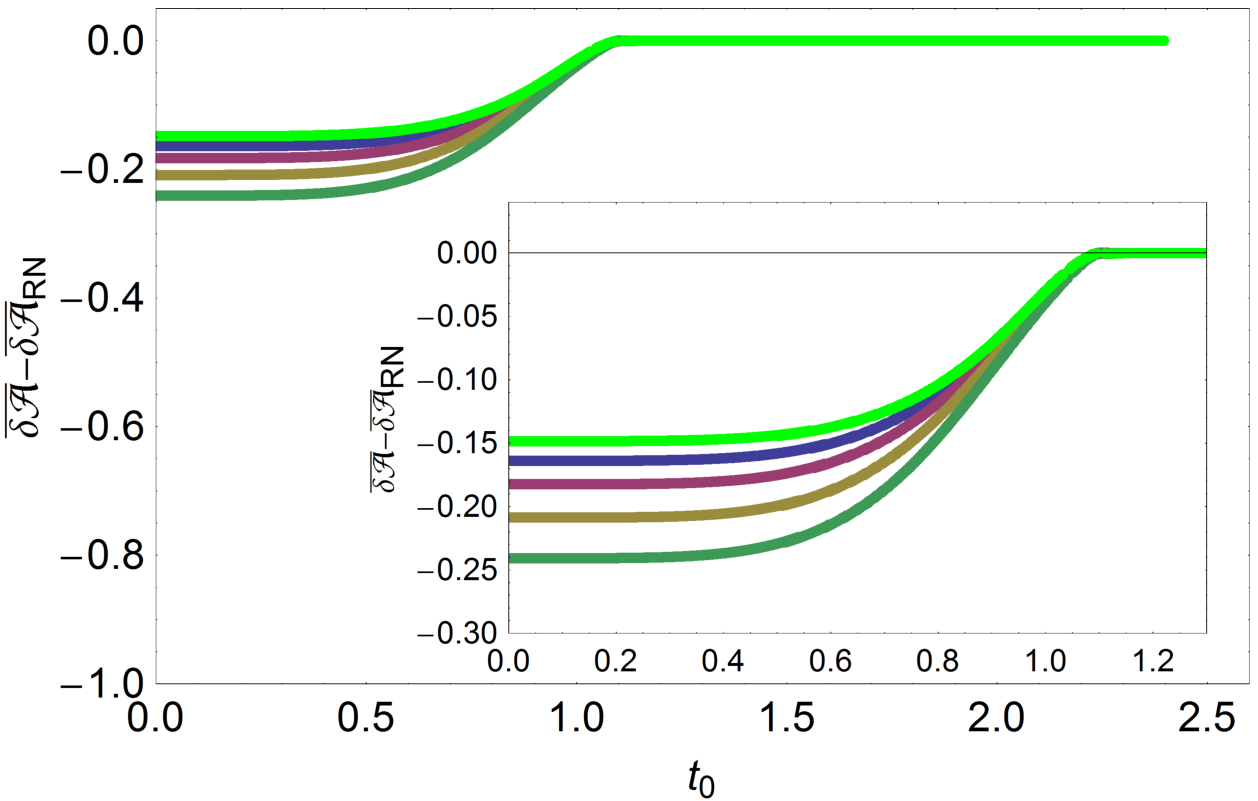}
\label{w15_ads7_RN} } \subfigure[AdS$_7$ $\&$ $\ell=2$]{
\includegraphics[scale=0.4]{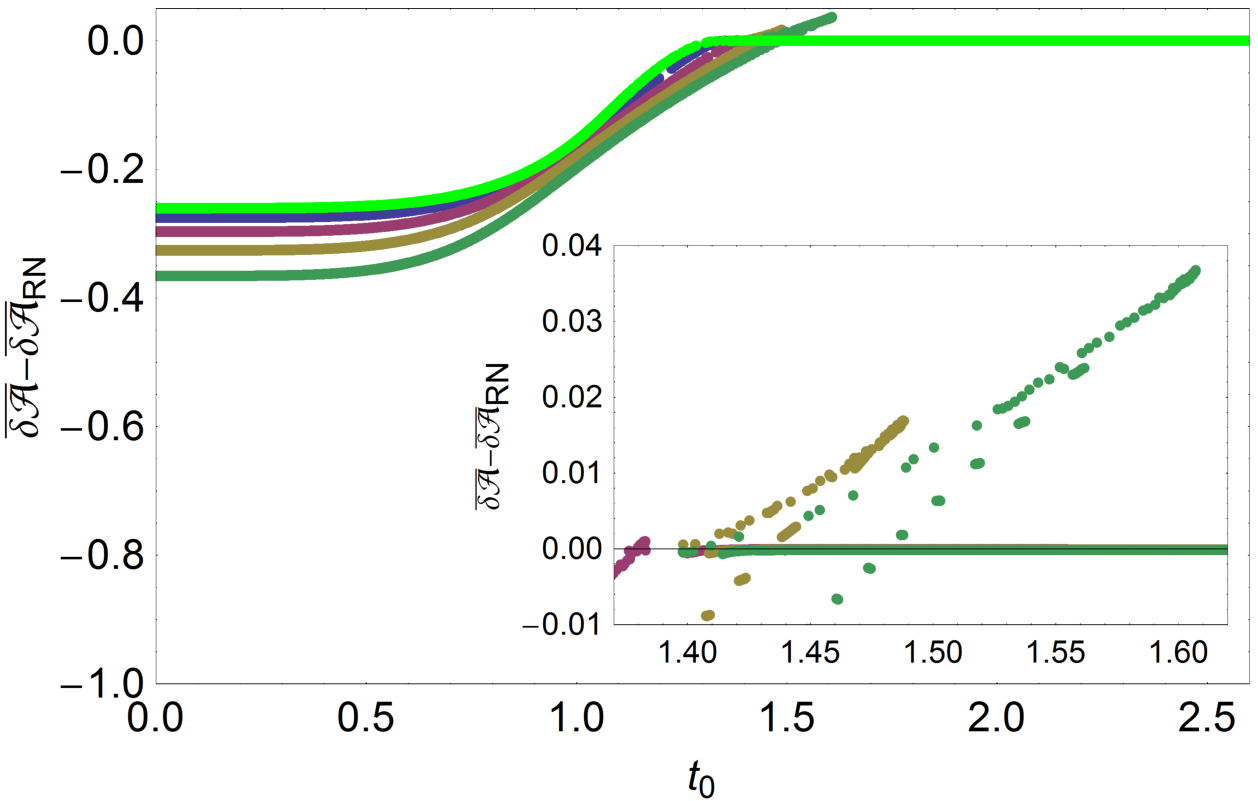}
\label{w2_ads7_RN} } \caption{\small Thermalizacion of the
renormalized minimal area surfaces for RNAdS Vaidya type metric with
$d+1=4, 5, 6, 7$, while the boundary separation is $\ell=1, 1.5, 2$.
In each picture the first curve (fluor green) indicates the case
$Q=0$, {\it i.e.} the Schwarzschild-AdS case. The rest of the curves
in each figure correspond to $Q=0.50, 0.75, \cdots,
\sqrt{\frac{d}{d-2}}$. The last case is the extremal where
$\mu/T\rightarrow\infty$. Insets allow to show the swallow tails in
detail. Note that in order to make easier the comparison among
different figures the vertical and horizontal scales are kept fixed
in all figures.} \label{fig wl RN}
\end{figure}

One of the most interesting effects we have found is the presence of
a swallow-tail pattern near the end of the thermalization process.
This phenomenon has been reported earlier in
\cite{Balasubramanian:2010ce} for thermalization of rectangular
Wilson loops in AdS$_4$ having $\mu=0$, as well as for those cases
when the shell is quasi-stationary. The reason for the emergence of
the swallow-tail structure is that, at certain time, there are more
than just only one single geodesic minimizing the action. This means
that the saddle-point approximation should be taken cautiously. In
reference \cite{Balasubramanian:2010ce}, on the other hand, it has
been argued that the swallow-tail structure depends on the dimension
of the system. However, according to our results displayed in Figs.
\ref{fig termalizacion tp RN} and \ref{fig wl RN}, this seems to be
a universal phenomenon which depends on the penetration of the
probes in the bulk interior. In fact, we show how for sufficiently
large probes swallow tails are independent on the kind of probe and
on the dimension of the asymptotic AdS space. Again, larger
dimensions induce smaller variations. This effect is understood by
considering that the fraction of the perturbed modes by the shell is
smaller as the dimensionality of the system increases.

\section{Discussion}

In this section we discuss the numerical results obtained by using
the formalism developed in the previous sections. We first analyze
the results of renormalized geodesic lengths, then minimal area
surfaces which are dual to expectation values of rectangular Wilson
loops and finally we discuss about the meaning of the swallow tails
in the figures. In all the figures corresponding to the geodesic
lengths and minimal area surfaces the curves correspond to values of
the charge from $Q=0$ to $Q_{ext}=\sqrt{\frac{d}{d-2}}$, with
increments of 0.25. Recall that the mass is given by the relation
$M=1+Q^2$. The UV cut-off has been taken as $z_0=0.01$, we have set
$r_h=1$ and the shell thickness is $v_0=0.01$.

Figure \ref{fig termalizacion tp RN} shows the results of
thermalization with a chemical potential for different spacetime
dimensions indicated as AdS$_{d+1}$, with $d+1=$ 4, 5, 6 and 7. In
all these cases we consider two-point functions as extensive probes.
As an example we show three different values of the boundary
separation $\ell=$2, 3 and 4. The pictures are displayed in such a
way that $\ell$ varies horizontally, while the spacetime dimension
changes in the vertical direction. In every picture the vertical
axis indicates the renormalized geodesic length as a function of
$t_0$. The fluor-green curve closer to the horizontal axis shows
that faster thermalization corresponds to the Schwarzschild-AdS
black hole. The rest of curves in all pictures indicates increasing
values of the $\mu/T$ ratio. Thus, they correspond to thermalization
with increasing chemical potential described in terms of
Reissner-Nordstr\"om AdS$_{d+1}$ black holes. The range of $\mu/T$
ratio goes from zero to infinity, and we show several curves where
always the fluor-green ones correspond to $\mu/T=0$ which coincide
with the evolution of thermalization reported in reference
\cite{Balasubramanian:2011ur}. As the chemical potential over
temperature ratio increases the curves depart from the fluor-green
ones, and the larger differences occur for curves corresponding to
$\mu/T \rightarrow \infty$. Looking at the first column of pictures
we observe that for small separation distances of local quantum
field theory operators at the boundary, {\it i.e.} $\ell=2$, the
renormalized geodesic length becomes zero at the same $t_0$. For
$\ell=3$ (in the second column of pictures) it is already possible
to notice a slight enhancement of the thermalization time as $\mu/T$
increases. This effect is much more noticeable for $\ell=4$, in the
third column of the pictures. Thus, the conclusion is an enhancement
of the thermalization time as the chemical potential over
temperature ratio increases, which is more evident as the boundary
separation increases.

Figure \ref{fig wl RN} shows the results of thermalization with a
chemical potential for different spacetime dimensions indicated as
AdS$_{d+1}$, with $d+1=$4, 5, 6 and 7, for minimal area surfaces
which turn out to be dual to expectation values of rectangular
Wilson loops. Notice that we display values of the boundary
separation, $\ell=$ 1, 1.5 and 2, which are different from the ones
shown in Fig. 2. From the pictures we can see that $\ell$ varies
horizontally, while the spacetime dimension changes in the vertical
direction as for the geodesic lengths. In the present case vertical
axis indicates the renormalized minimal area surfaces as functions
of $t_0$. The fluor-green curve closer to the horizontal axis shows
that faster thermalization corresponds to the Schwarzschild-AdS
black hole as before. The curves correspond to thermalization with
increasing chemical potential described in terms of
Reissner-Nordstr\"om AdS$_{d+1}$ black holes. The range of $\mu/T$
ratio goes from zero to infinity, and we show several curves where
always the fluor-green ones correspond to $\mu/T=0$. As the chemical
potential increases the curves depart from the fluor-green one, and
the larger differences occur for curves corresponding to $\mu/T
\rightarrow \infty$. Thermalization time is obtained when the curves
reach zero. For $\ell=1.5$ (in the second column of pictures) it is
already possible to notice a slight enhancement of the
thermalization time as $\mu/T$ increases. This effect is more
noticeable for $\ell=2$. Therefore, we arrive to the same conclusion
as in the geodesic lengths case: there is an enhancement of the
thermalization time as the chemical potential over temperature ratio
increases, which is more evident as the boundary separation
increases.

Each figure \ref{fig termalizacion tp RN}(a-l) as well as \ref{fig
wl RN}(a-l) shows that the thermalization time for renormalized
geodesic lengths and minimal area surfaces increases as $\mu/T$
increases, while the dimension of the bulk spacetime $d+1$ and the
boundary separation $\ell$ of two local field theory operators are
kept fixed. When considering the three figures for a fixed dimension
in each horizontal sequence, there are two effects: one is that the
thermalization time increases as the boundary separation does; the
second effect is that for small $\ell$ there are no apparent
differences in the thermalization time for any value of $\mu/T$ and,
moreover, it coincides with the Schwarzschild-AdS black hole case.
As long as one increases $\ell$ (see figures \ref{fig termalizacion
tp RN} and \ref{fig wl RN} (c, f, i, l)), the thermalization is
significantly delayed as $\mu/T$ increases. This effect is
understood as a consequence of the fact that the term proportional
to the charge in $f_{RN}(v,z)$ has opposite relative sign with
respect to the term proportional to the mass, so the chemical
potential somehow produces an effect similar to the one obtained
from the reduction of the black hole mass. In addition, it turns out
that the thermalization time is slightly larger for expectation
values of rectangular Wilson loops compared with two-point
functions. For the same separation length in the boundary, the
minimal area surface dual to the expectation values of Wilson loops
penetrate more into the bulk, thus resulting in a more extensive
probe of thermalization, which takes more time to measure thermal
equilibrium. From the dual quantum field theory side, we expect that
the system after receiving a sudden injection of energy and
particles will be far from thermal equilibrium. We assume that its
dynamics should be controlled by a strongly coupled Yang Mills
theory. The study of the thermalization process cannot be done in
terms of the hydrodynamic approximation, and therefore, the
description of the process is a very hard problem. On the other
hand, we can think of a rather intuitive argument, which although
has obviously not the value of any rigorous proof, it may help to
understand whether our results obtained using a holographic dual
model of thermalization are qualitatively consistent with the gauge
theory expectations. First notice that in order for the system to
reach thermal equilibrium it requires both kinetic and chemical
equilibration. The former is expected to be faster than the second
one. The reason is that chemical equilibration involves additional
interactions. This, in principle, might delay the thermalization
process in comparison with the situation where no chemical potential
is considered.

Also from these figures a swallow-tail pattern emerges. In the
insets of the figures we zoom in the corresponding regions of the
figures in order to see the swallow-tail structure in more detail.
Now, let us consider the figures for a fixed $\ell$. So, increasing
the AdS dimension we observe that the effects of chemical potential
become less relevant in all cases. This is due to the fact that a
geodesic is a one-dimensional line which is not able to account for
the effect of other orthogonal space dimensions. Thus, if we
increase $d+1$, in order to observe the effect on the thermalization
time, we have to use an extended probe having more space dimensions.
This effect can be seen by comparison with the thermalization curves
using the dual of expectation values of Wilson loops as extended
probes. In all the figures we can observe that UV degrees of freedom
thermalize first.

\begin{figure}
\centering \subfigure[$z_*=1.207$ $\&$ $v_*=-0.194$]{
\includegraphics[scale=0.4]{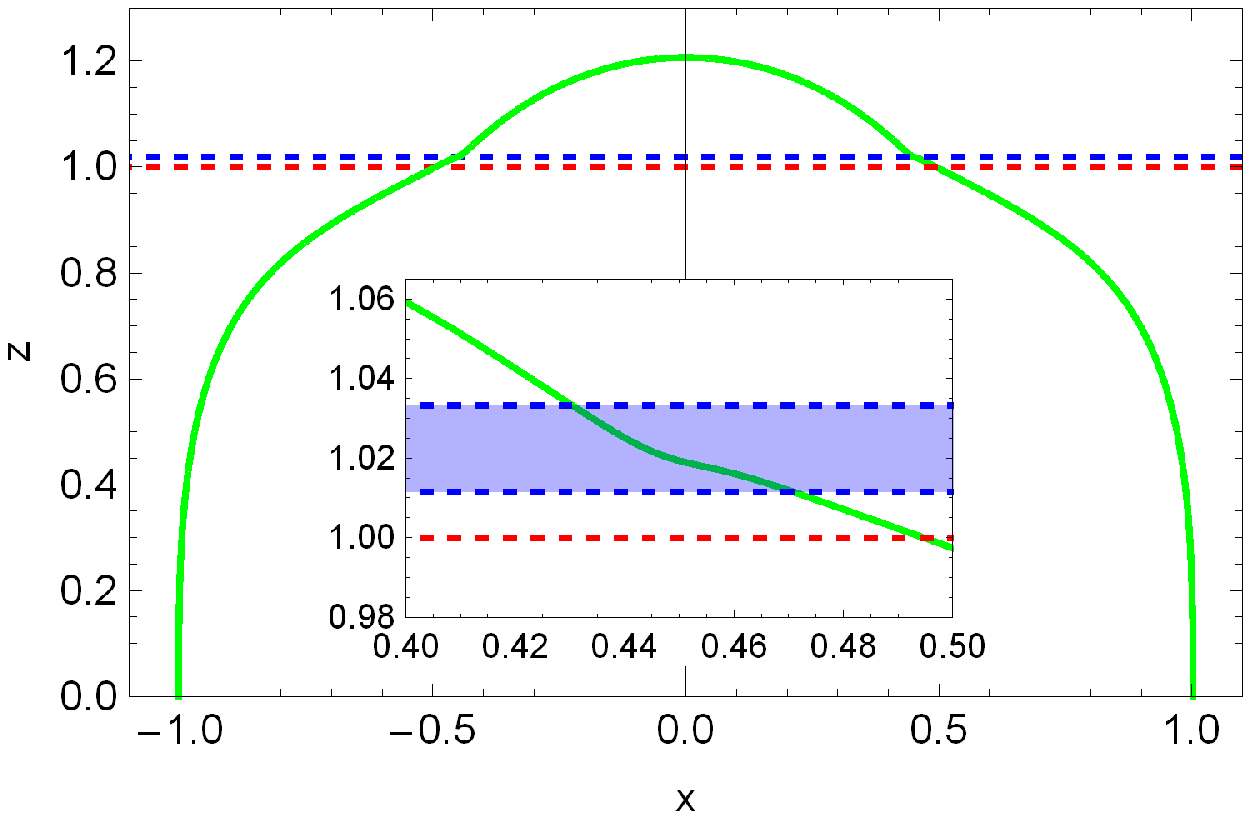}
\label{swallow1} } \subfigure[$z_*=1.201$ $\&$ $v_*=-0.189$]{
\includegraphics[scale=0.4]{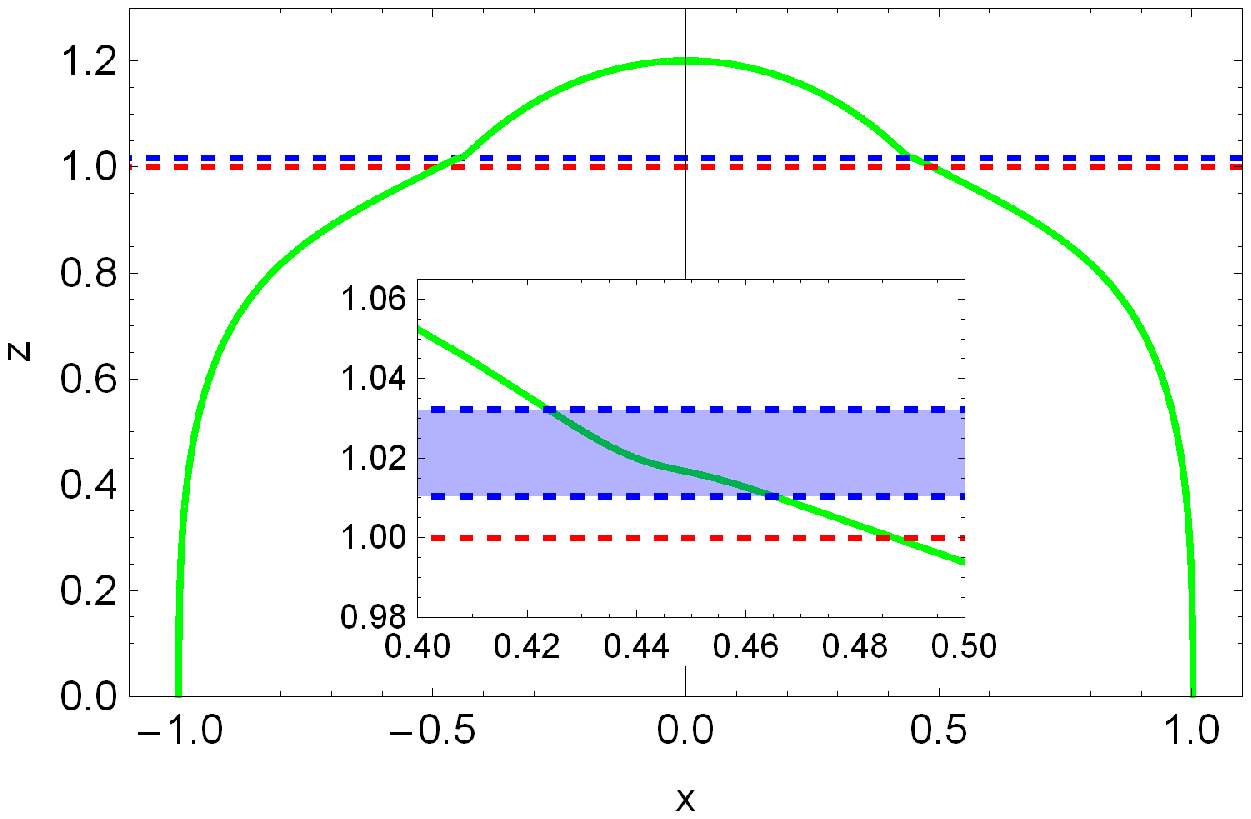}
\label{swallow2} } \subfigure[$z_*=0.839$ $\&$ $v_*=0.173$]{
\includegraphics[scale=0.4]{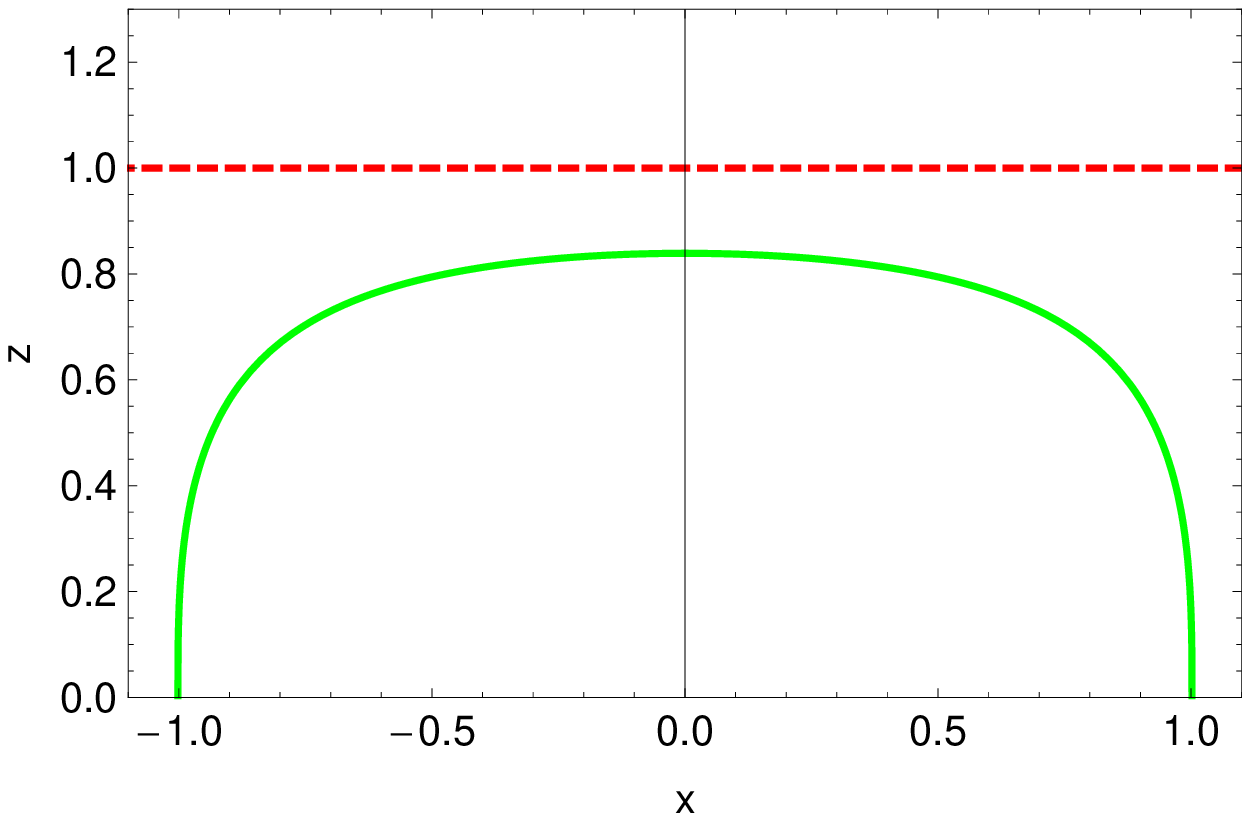}
\label{swallow3} } \caption{\small Three different solutions that
appear as part of the swallow tail in thermalization of expectation
values of Wilson loops in the extremal RNAdS Vaidya geometry with
$d=3$. Time corresponds to $t_0=1.778$ in all three cases and
separation length is $\ell=2$. The red dashed line corresponds to
the event horizon ($z_h=1$), while the blue dashed one is the
position of the shell. In the first two cases, the shell is located
inside the event horizon but still have solutions inside. In the
insets we take a closer look at the region near the horizon and
consider the shell thickness ($-0.02\leq v \leq 0.02$) in blue.}
\label{swallow}
\end{figure}

It is interesting to analyze the cause of the swallow-tail emergence
in thermalization curves with large boundary separation distance. In
general this kind of effect is found in systems where there are two
different scales. In the present case, there are different length
scales that could, in principle, induce the swallow-tail pattern.
For instance, the AdS radius, the inverse of confinement temperature
$T_c$, the UV cut-off $z_0$ and the shell thickness, can all
introduce a second scale to the problem. However, we do not find
that any of these scales are relevant for the appearance of a
swallow tail. In fact, we will show that different solutions given
in the swallow tails are actually non-physical solutions of our
problem, in the sense that they have not to be taken into account.
This can be shown by looking at how the solutions that appear in the
swallow tail are. As an example, we show in Fig. \ref{swallow} these
solutions for the swallow tail that appear when evaluating
expectation values of rectangular Wilson loops in the extremal
RNAdS$_4$ Vaidya metric. What we can appreciate is that from the
three different solutions that generate the swallow tail at a
certain time, two of them have $z(x)$ functions that propagate
inside the event horizon. One could nevertheless think that the
appearance of those solutions is due to the shell thickness that can
emulate a black hole solution with a deeper event horizon. However,
as it is shown in the insets of Figs. \ref{swallow1} and
\ref{swallow2} even by considering the shell thickness we obtain
that these solutions propagate beyond the event horizon. So, only
the third one, that appears in Fig. \ref{swallow3}, is the physical
and thermalized solution.

The effective model presented above is based on a double quench (for
the mass and charge) and it has a metric which evolves in time.
These features make, in principle, difficult to think of finding a
top-down description from string theory and M-theory. However, it is
certainly interesting to study this kind of models based on a
collapsing shell of charged dust since it allows to study the
thermalization process in a dual strongly coupled quantum field
theory. One of the important results is that by using this geometric
extended probes one finds a top-down thermalization regardless of
the dimension of the AdS space and of whether the shell is charged
or not. This is a consequence of the geometric setup, where for
smaller boundary separations the dual geometric structures probe a
region closer to the boundary. Also, as it is the case for
Schwarzschild-AdS black holes there is a delay in the beginning of
the thermalization process due to the fact that the system only
feels the sudden energy injection at the boundary at distances of
its thermal wavelength. We have carried out a systematic
investigation by exploring the whole range of $\mu/T$, the boundary
separation and the dimension of the AdS spaces. Except for the case
of the BTZ black hole the expressions require to solve them
numerically. The reason to include chemical potential, using the
grand canonical ensemble, is because realistic systems such as QGPs
and condensed matter systems have that. In order to do so, the
formalism presented in this work, studying a general
Reissner-Nordstr\"om-AdS black hole Vaidya-type metric in arbitrary
number of dimensions, is very interesting and we hope it will
motivate further studies and extensions. In all the cases studied
the effect produced by the chemical potential is the delay of the
thermalization process, being it more evident as the boundary
separation increases. This also agrees with the conclusion that the
thermalization of more energetic modes proceeds first in strongly
coupled systems.

~

\centerline{\large{\bf Acknowledgments}}

~

We thank Esteban Calzetta, Nicol\'as Grandi, Carlos N\'u\~nez and
Guillermo Silva for useful discussions and Carlos N\'u\~nez for a
careful reading of the manuscript. This work has been partially
supported by the CONICET, the ANPCyT-FONCyT Grant PICT-2007-00849,
and the PIP-2010-0396 Grant.

\newpage

\appendix
\section{On null vectors and energy conditions of the RN-AdS-Vaidya type metric}

In this Appendix we show how the energy conditions are satisfied for
a RN-AdS-Vaidya type metric like the one that has been analyzed in
this work as a gravitational dual to thermalization of field
theories at finite temperature with a finite chemical potential.

The energy-momentum tensor $T_{\mu\nu}$ at each point of a manifold
satisfies the so-called weak-energy condition \cite{Hawking:1973uf},
$T_{\mu\nu} W^\mu W^\nu \geq 0$ for any time-like vector $W^\mu$. In
particular, for any null vector $N^\mu$, this becomes the condition
$T_{\mu\nu} N^\mu N^\nu \geq 0$. In the case of the Vaidya metric we
will follow \cite{AbajoArrastia:2010yt} and consider normal null
vectors to surfaces that preserve translation invariance in the
boundary coordinates $x_i$; that is, surfaces of constant $v$ and
$z$. The resulting null vectors are
\begin{eqnarray}
N_1^\mu & = & \left( 0, \frac{z^2}{R^4}, \bf{0} \right), \\
N_2^\mu & = & \left( R^2, -\frac{f_{RN}(v, z)}{2}, \bf{0} \right) \,
,
\end{eqnarray}
where we have restored units of AdS radius $R$. With this choice the
metric and $f_{RN}$ becomes
\begin{eqnarray}
ds^2 & = & \frac{1}{z^2} \left[ - f_{RN}(v, z) dv^2 - 2 R^2 dz dv + R^2 d\bf{x}^2 \right] \\
f_{RN}(v, z) & = & R^2 - m(v) \frac{z^d}{R^{2d-4}} + q(v)^2
\frac{z^{2d-2}}{R^{4d-8}} \, .
\end{eqnarray}
The normalization has been chosen such that $N_1 \cdot N_2 = -1$. As
the only non-vanishing component of the energy-momentum tensor is
\begin{eqnarray}
8 \pi G^{d+1} T_{vv}= \frac{(d-1) z^{d-1}}{R^{4d-6}} \left(
\frac{1}{2} R^{2d-4} \dot{m}(v) - z^{d-2} q(v) \dot{q} (v) \right)
\, ,
\end{eqnarray}
the null energy condition for $N_1$ is satisfied trivially. For
$N_2$ the energy condition depends on the relation between $m(v)$
and $q(v)$. As mentioned before, for $q(v)=0$ the condition is
always satisfied provided that $\dot{m}(v)$ is non-negative. In the
charged case, the expression can be analyzed in a simpler way by
making use of the relation between mass, charge and event horizon
radius given in Eq.(\ref{masa RN}). The chosen quenches satisfy
\begin{eqnarray}
m(v) &=& \frac{M}{2} g(v) \, , \\
q(v) &=& \frac{Q}{2} g(v) \, ,
\end{eqnarray}
where $g(v)=2 \theta(v)$ in the zero shell thickness case and
$g(v)=1+\tanh(\frac{v}{v_0})$ in the general case. The null energy
condition becomes
\begin{eqnarray}
 \frac{d-1}{4} \frac{z^{d-1}}{R^{4d-10}} \dot{g}(v) \left(
\frac{R^{4d-6}}{z_h^{d}}  + Q^2 \left(z_h^{d-2}-z^{d-2}g(v)
\right)\right)\geq 0  \, .
\end{eqnarray}

The overall function $\dot{g}(v)$ makes that the condition is
satisfied if
\begin{eqnarray}
z \leq \frac{1}{g(v)^{\frac{1}{d-2}}} \, \left( \frac{z_h^{-d}
R^{4d-6}}{Q^2}+ z_h^{d-2} \right)^{\frac{1}{d-2}} \, .
\end{eqnarray}
For the zero shell thickness, i.e. for the step function, the null
energy condition is satisfied for any value of $v$. In the case of a
smoother function, there can be a small region for large $Q$ around
$v=0$, where the condition is not satisfied. On the other hand, we
can always consider $R$ and $z_h$ to be within a parametric region
where $R^{4d-6}/z_h^{2d-2} \geq Q^2$, where the null energy
condition is satisfied for any value of $v$. In addition, as the
limit case of zero shell thickness is indeed physical, we consider
that it is interesting to study the whole range of parameters from
zero charge to the extremal case.

\end{document}